\documentclass[journal]{IEEEtran}  %

\usepackage{lscape}

\usepackage{multirow}     %
\usepackage{booktabs}     %
\usepackage{amsmath,amssymb,amsfonts}
\usepackage{rotating}
\usepackage{graphicx}
\usepackage{algorithm}
\usepackage{algorithmic}
\usepackage{adjustbox}
\usepackage{tabularray}
\usepackage[graphicx]{realboxes}
\usepackage{diagbox, subfigure}
\usepackage{cite}

\begin{document}

\bstctlcite{IEEEexample:BSTcontrol}
\title{Low Power Vision Transformer Accelerator with Hardware-Aware Pruning and Optimized Dataflow
}

\author{\IEEEauthorblockN{Ching-Lin Hsiung, and Tian-Sheuan Chang, \textit{Senior Member, IEEE}}
\thanks{This work was supported by the National Science and Technology Council, Taiwan, under Grant 111-2622-8-A49-018-SB, 110-2221-E-A49-148-MY3, 113-2221-E-A49-078-MY3, and 113-2640-E-A49-005. The authors are affiliated with the Institute of Electronics, National Yang Ming Chiao Tung University, Taiwan. (e-mail: bear.11@nycu.edu.tw, tschang@nycu.edu.tw) }%
\thanks{Manuscript received XXXX XX, 2025; revised XXXX XX, XXXX.}
}
\maketitle

\begin{abstract}%
Current transformer accelerators primarily focus on optimizing self-attention due to its quadratic complexity. However, this focus is less relevant for vision transformers with short token lengths, where the Feed-Forward Network (FFN) tends to be the dominant computational bottleneck. This paper presents a low power Vision Transformer accelerator, optimized through algorithm-hardware co-design. The model complexity is reduced using hardware-friendly dynamic token pruning without introducing complex mechanisms. Sparsity is further improved by replacing GELU with ReLU activations and employing dynamic FFN2 pruning, achieving a 61.5\% reduction in operations and a 59.3\% reduction in FFN2 weights, with an accuracy loss of less than 2\%. The hardware adopts a row-wise dataflow with output-oriented data access to eliminate data transposition, and supports dynamic operations with minimal area overhead. Implemented in TSMC's 28nm CMOS technology, our design occupies 496.4K gates and includes a 232KB SRAM buffer, achieving a peak throughput of 1024 GOPS at 1GHz, with an energy efficiency of 2.31 TOPS/W and an area efficiency of 858.61 GOPS/mm².

~
\end{abstract}

\section{Introduction}
\label{chapter:Related Work}

Transformers have gained significant traction in vision applications due to their superior feature extraction capabilities and scalability. However, their widespread adoption in embedded and low-power systems is hindered by their high computational complexity, memory bandwidth demands, and energy inefficiency. Vision Transformers (ViTs), in particular, require specialized hardware accelerators to achieve real-time inference while maintaining power and area efficiency. Unlike traditional AI accelerators designed for general-purpose computation, hardware-efficient ViT accelerators must incorporate power-aware architectural optimizations and customized dataflow strategies to improve performance while reducing energy consumption and area overhead.

\begin{figure}[htbp]
\centering
\includegraphics[width=0.8\linewidth, keepaspectratio=true]{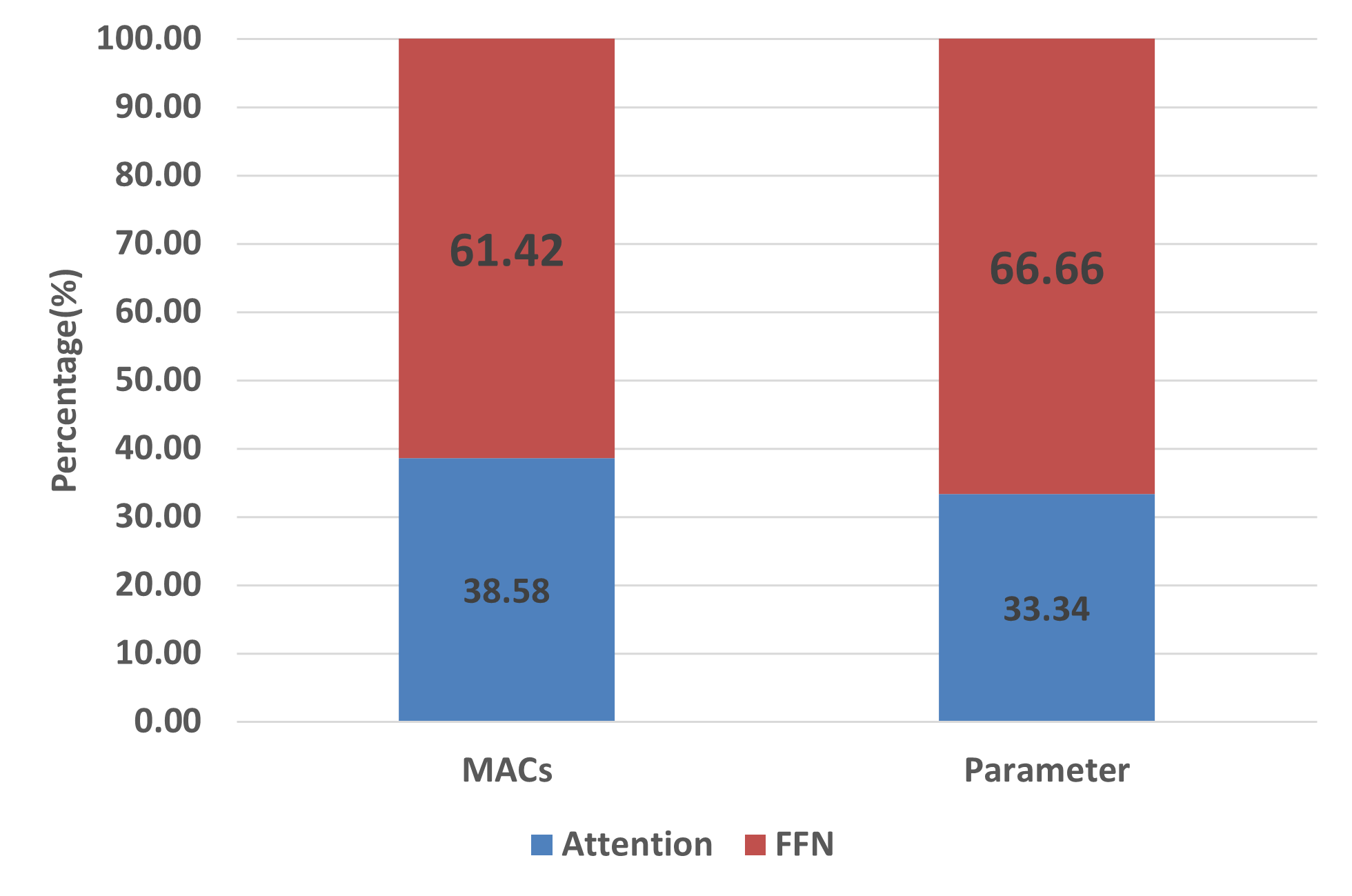}
\caption{Distribution analysis of the DeiT-small model}
\label{ViT_Analysis}
\end{figure}

For the hardware design of transformers, 
while the quadratic complexity of self-attention is a well-recognized challenge, in ViTs commonly used for vision tasks (e.g., DeiT-Small~\cite{DeiT}) that process a moderate number of tokens (e.g., 197 for $224\times224$ images), the Feed-Forward Networks (FFNs) often account for over 60\% of the total MAC operations (as shown in Fig.~\ref{ViT_Analysis}) and a significant portion of parameter storage. 
Therefore, efficient ViT accelerator design must address not only attention mechanisms but also, and perhaps more critically, the computational load and memory footprint of FFNs. 
Transformer-specific accelerators such as $A^3$~\cite{A3}, ELSA~\cite{ELSA}, and~\cite{28nm} focus on optimizing self-attention through methods like approximate computations, but they do not address FFN optimization. Meanwhile, approaches like SpAtten~\cite{SpAtten} and FACT~\cite{FACT} incorporate token pruning, with FACT extending its optimization to both the attention mechanism and FFN, using dynamic pruning informed by early correlation predictions. However, both SpAtten and FACT, while effective in reducing computational load, still leave room for further optimization, especially in terms of FFN pruning and computational overhead management.

Additional techniques such as quantization and dynamic token pruning have also been explored to accelerate computation. For example, 
\cite{wang2024bsvit} uses bit-serial computation with dynamic patch quantization to improve energy efficiency. 
Dynamic token pruning has become a popular research area in Transformer model optimization to introduce sparsity and save the computational demands of self-attention and FFNs, both of which scale with the number of tokens. Dynamic token pruning consists of two types: parameter-free pruning~\cite{Evo-ViT, EViT, A-ViT, ATS} and pruning with additional prediction submodules~\cite{DynamicViT, SPViT, HeatViT}. In each type, the pruning rate can be static~\cite{DynamicViT, SPViT, Evo-ViT} or dynamic~\cite{HeatViT, A-ViT, ATS}.  However, many of these approaches are not optimized for hardware efficiency and introduce substantial control overhead, limiting their practicality in real-world hardware deployments. In summary, ViT accelerators targeting edge applications should exploit sparsity in a hardware-efficient and low-overhead manner to reduce computational complexity and memory access, not only in self-attention but also in the FFN modules. Furthermore, the design challenges of ViT accelerators extend beyond sparsity and pruning; they must also address the inefficiency of hardware-unfriendly activation functions and optimize dataflow to enhance hardware utilization and reduce unnecessary data movement.

Addressing the above design challenges, this paper presents a low-power Vision Transformer accelerator through hardware-software co-design that integrates hardware-aware optimizations to reduce computational complexity and improve system-level efficiency. Our proposed accelerator features:
\begin{itemize}
    \item Dynamic Token Pruning: A lightweight hardware-compatible pruning strategy that reduces computation without requiring complex control mechanisms.
    \item FFN2 Pruning: A memory-efficient pruning approach that reduces unnecessary weight operations, leading to lower power consumption and area overhead.
    \item Hardware-Friendly Activation Function: By replacing GELU with ReLU, we improve activation sparsity and simplify the circuit design for activation computations.
    \item Optimized Dataflow Architecture: The accelerator employs a row-wise, output-oriented dataflow, minimizing data transposition overhead and reducing memory bandwidth requirements.
\end{itemize}

Implemented in TSMC’s 28nm CMOS technology, our accelerator achieves a peak throughput of 1024 GOPS at 1GHz, with an energy efficiency of 2.31 TOPS/W and an area efficiency of 858.61 GOPS/mm². Compared to existing transformer accelerators, our design achieves a 61.5\% reduction in computations, a 59.3\% reduction in FFN2 weights, and a 56.4\% reduction in external memory access, making it highly suitable for power- and area-constrained applications in embedded vision systems.

The rest of the paper is organized as follows: Section II details the proposed model optimizations, Section III describes the hardware architecture, Section IV presents the experimental results, and Section V concludes the paper.

\section{Proposed Method}
\begin{figure}[htbp]
\centering
\includegraphics[width=0.8\linewidth, keepaspectratio=true]{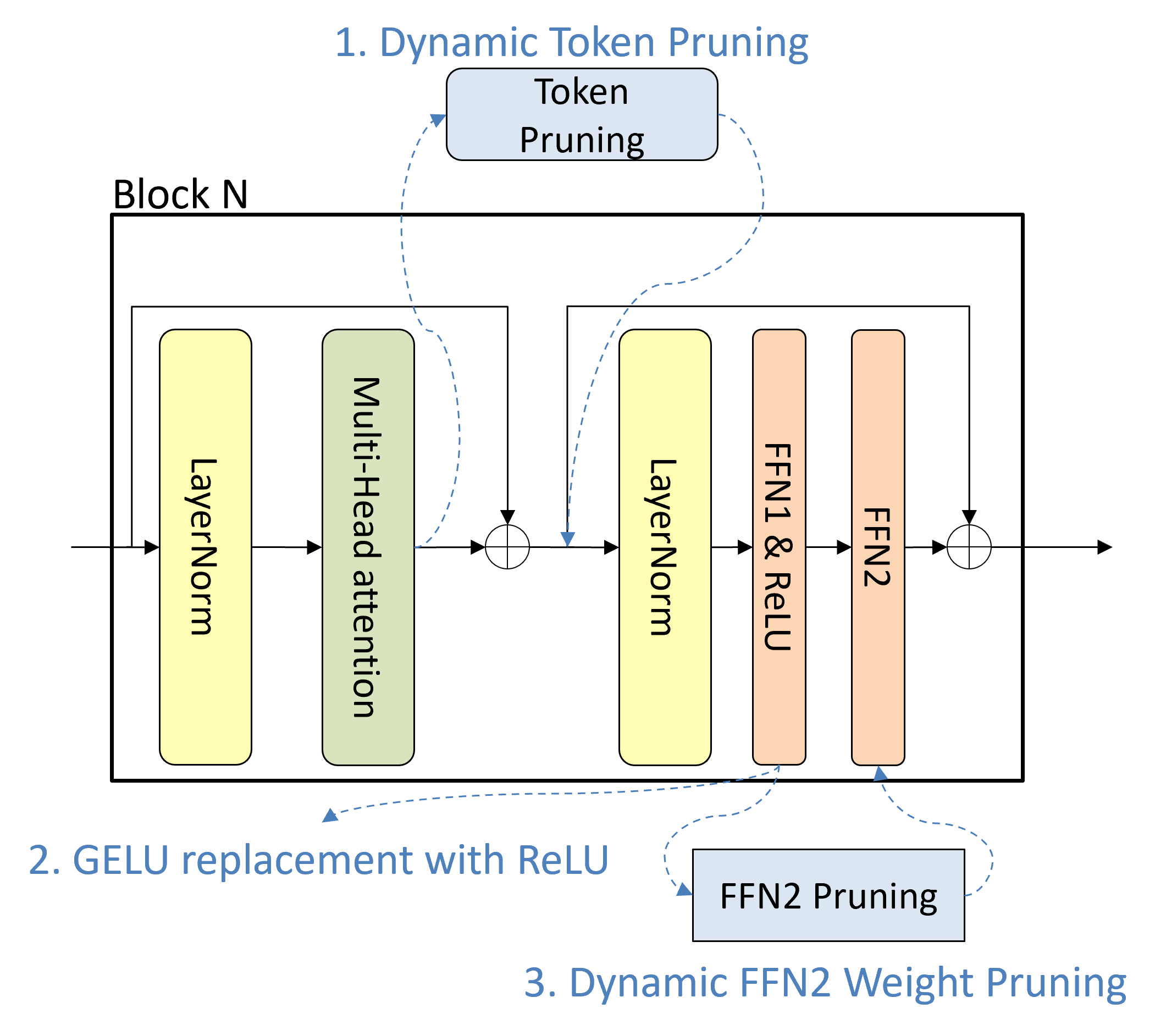}
\caption{Inference process within a single encoder block}
\label{Method_overview}
\end{figure}

To efficiently implement ViT acceleration in hardware, we propose a hardware-aware methodology that optimizes computational efficiency, memory access patterns, and power consumption. Our approach incorporates hardware-friendly dynamic token pruning, low-cost activation functions, and dynamic FFN2 weight pruning, as shown in Fig.~\ref{Method_overview}, ensuring a balance between accuracy and circuit efficiency.

\subsection{Hardware-Friendly Dynamic Token Pruning}
\label{Dynamic Token Pruning}
Token pruning is widely used to reduce computational complexity in transformers. However, existing token pruning techniques often introduce excessive control overhead, making them inefficient for hardware implementations. In our design, we implement a lightweight, hardware-friendly token pruning mechanism that minimizes unnecessary computations while maintaining a simple control flow.

\begin{figure}[htbp]
\centering
\includegraphics[width=0.8\linewidth, keepaspectratio=true]{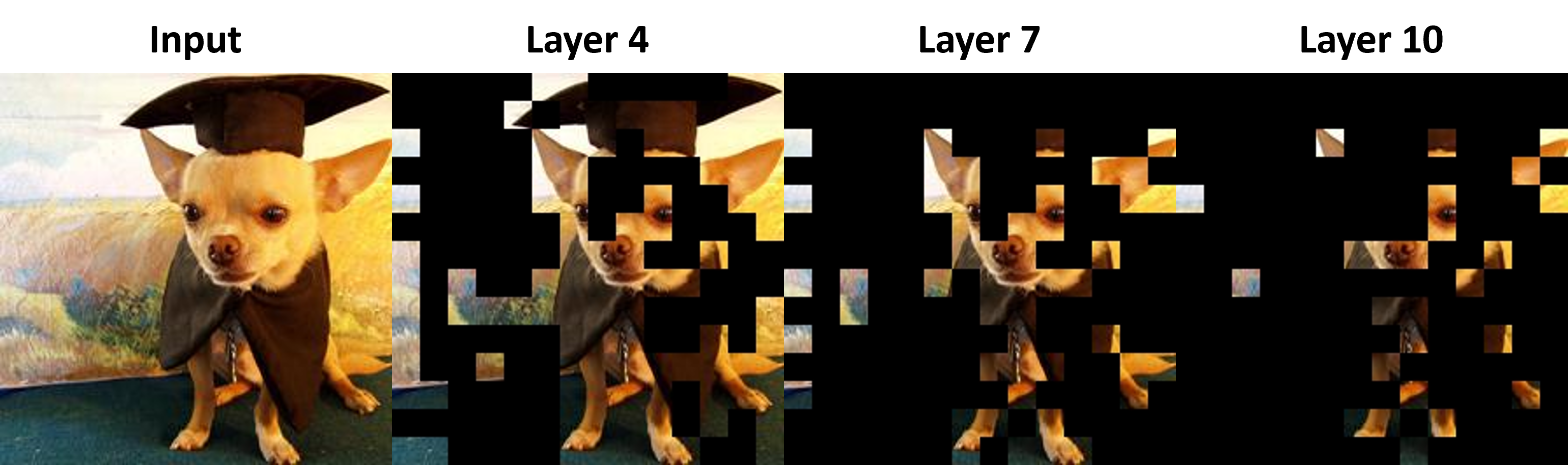}
\caption{Visualization of the token pruning process}
\label{Visualization}
\end{figure}

Token pruning removes unimportant tokens. In ViTs, an image is divided into patches and encoded as tokens for model input. These tokens vary in importance. In classification tasks, essential objects in the foreground contribute more to the decision-making process, while background elements and less relevant regions can often be discarded without significantly affecting accuracy. Fig.~\ref{Visualization} illustrates token pruning at Layers 4, 7, and 10, where irrelevant image regions are progressively pruned, allowing the model to focus on essential features, such as the dog in this example. This process reduces computational load while maintaining accuracy and efficiency.

The proposed method, inspired by ~\cite{EViT}, determines token importance using the class token's attention values. The attention vector for the class token in a single attention head is given by:

\begin{equation}
{\bf A}_h^{(cls)} = \text{SoftMax}\left(\frac{q_{cls} \cdot {\bf K}^T}{\sqrt{d}}\right) \cdot {\bf V} = {\bf a}_h^{(cls)} \cdot {\bf V}
\end{equation}

where ${\bf A}_h^{(cls)}$ represents the attention output for the class token in the $h$-th attention head, $q_{cls}$ is the query vector for the class token, and ${\bf K}$ denotes the key matrix for all tokens. The key dimension is denoted by $d$. The final class attention vector is computed as the average across all heads:

\begin{equation}
{\bf a}^{(cls)} = \frac{1}{H} \sum_{h=1}^{H}{\bf a}_h^{(cls)}
\end{equation}

Next, these attention values are sorted, and the Top-K values are selected to retain the most relevant tokens. Unlike ~\cite{EViT}, which merges less important tokens into one, the proposed method discards tokens outside the Top-K selection, avoiding unnecessary computation for fusion. Compared to prior works that rely on additional gating networks \cite{DynamicViT, SPViT}, our approach leverages a threshold-based selection mechanism. This method enables real-time token selection without requiring complex learning-based mechanisms. 

\subsection{Hardware-Friendly Activation}
\label{Hardware-Friendly Activation}

Gaussian Error Linear Unit (GELU) is a widely used non-monotonic activation function in ViTs~\cite{ViT, DeiT, liu2021swin, yuan2021tokens, touvron2021going}, approximated as:

\begin{equation}
\text{GELU}(x) \approx 0.5 \cdot x \left[ 1 + \tanh \left( \sqrt{\frac{2}{\pi}} \left( x + 0.044715x^3 \right) \right) \right]
\end{equation}

However, GELU is computationally complex due to its reliance on the error function and precise approximations, making it less suitable for hardware implementations. This limitation necessitates estimation techniques or lookup tables, increasing computational overhead and potential accuracy degradation.

To address this, we replace GELU with ReLU, simplifying the activation process and enhancing sparsity by eliminating negative values. This change improves computational efficiency and aligns better with hardware constraints.

Our findings show that replacing GELU with ReLU and training from scratch for 300 epochs stabilizes post-activation values and increases sparsity across all layers. However, pre-activation values exhibit significant accumulation of small negative values, and training from scratch with ReLU results in noticeable accuracy degradation. To mitigate this, we propose a finetuning approach using pretrained models originally trained with GELU. Two strategies are adopted:

\begin{enumerate}
\item \textit{Full Replacement and Finetuning:} Replace all GELU activations with ReLU and finetune for 35 epochs. This approach achieves accuracy levels close to the original GELU-based model.
\item \textit{Layer-by-Layer Replacement and Finetuning:} Replace GELU with ReLU incrementally, finetuning for 2-3 epochs per layer over a total of 35 epochs. This approach outperforms full replacement, achieving accuracy even closer to the original model.
\end{enumerate}

\begin{figure}[htbp]
\centering
\includegraphics[width=1.0\linewidth, keepaspectratio=true]{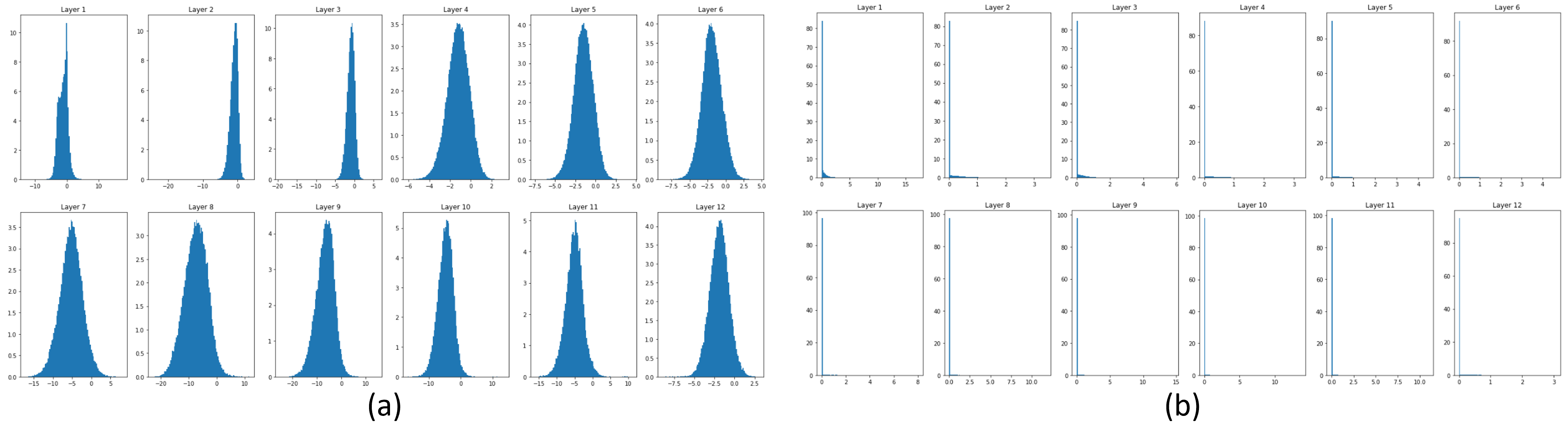}
\caption{Histogram of FFN intermediate results before and after ReLU finetuning. (a) Before ReLU (b) After ReLU}
\label{ReLU_finetune}
\end{figure}

With the second method, Fig.~\ref{ReLU_finetune} demonstrates that post-activation values maintain high sparsity and stability, while pre-activation values become more normally distributed. This technique effectively combines ReLU's hardware efficiency with the stability and accuracy associated with GELU.

\subsection{Dynamic FFN2 Weight Pruning}
\label{Dynamic FFN2 Weight Pruning}

\subsubsection{Observation}
\begin{figure}[htbp]
\centering
\includegraphics[width=0.45\linewidth, keepaspectratio=true]{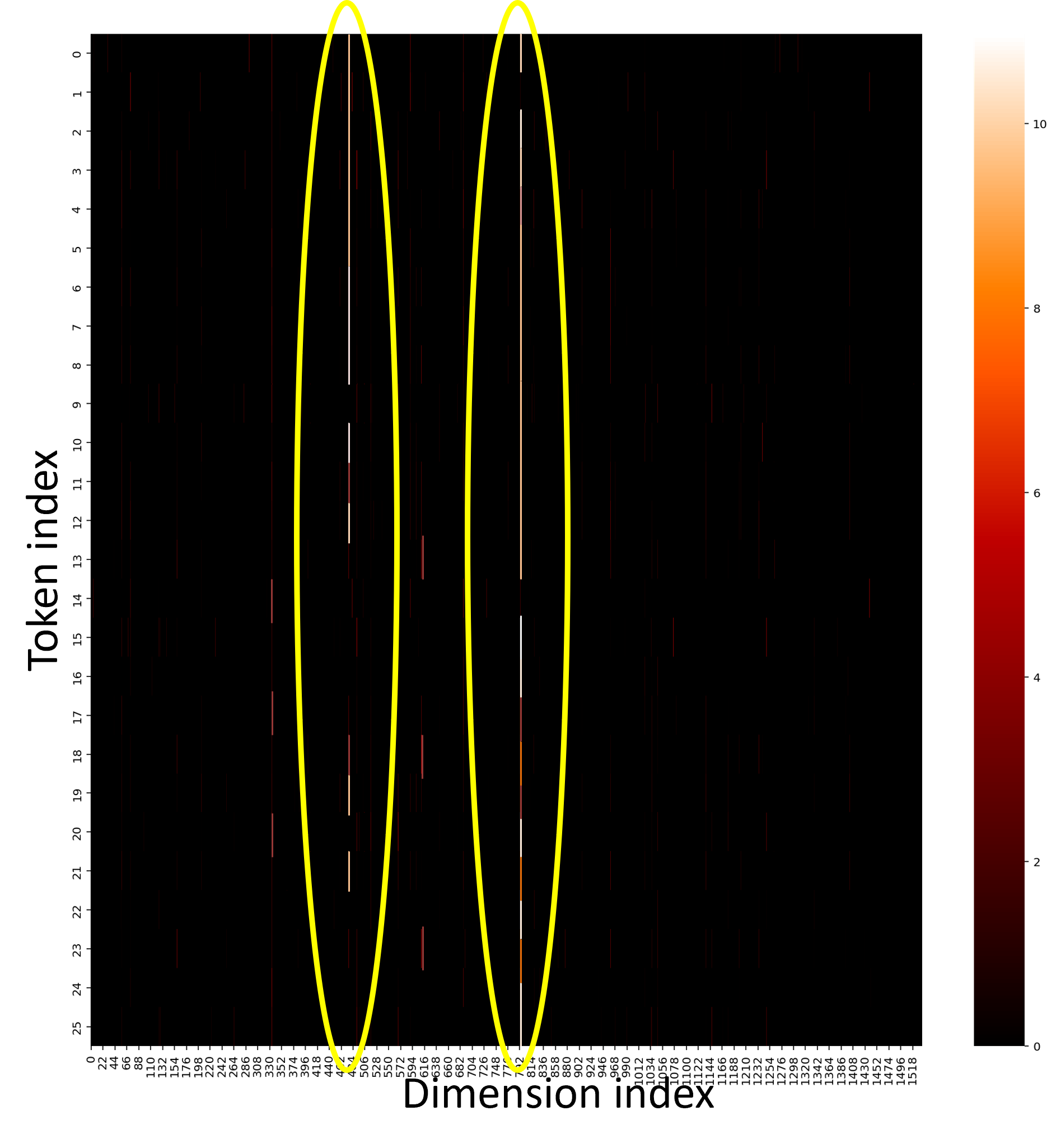}
\caption{Distribution of FFN post-activation matrix values}
\label{distribution}
\end{figure}

The FFN in ViTs constitutes a major computational bottleneck, consuming significant power and memory bandwidth. To alleviate this, we introduce FFN2 pruning, which selectively removes low-importance activations to optimize both computation and storage requirements. 
Our FFN2 pruning mechanism integrates the ReLU-based activation sparsity detector that identifies redundant neuron activations before weight multiplication. Fig.~\ref{distribution} shows the value distribution of the FFN post-activation matrix in deeper layers. Most values are either zero or small positive numbers (dark areas), indicating high sparsity. However, non-zero values are concentrated in specific dimensions (highlighted in yellow), suggesting that only a few dimensions carry significant information, while the rest contribute minimally. This motivates a dynamic pruning strategy focusing on key dimensions to optimize efficiency.

\begin{figure}[htbp]
\centering
\includegraphics[width=1.0\linewidth, keepaspectratio=true]{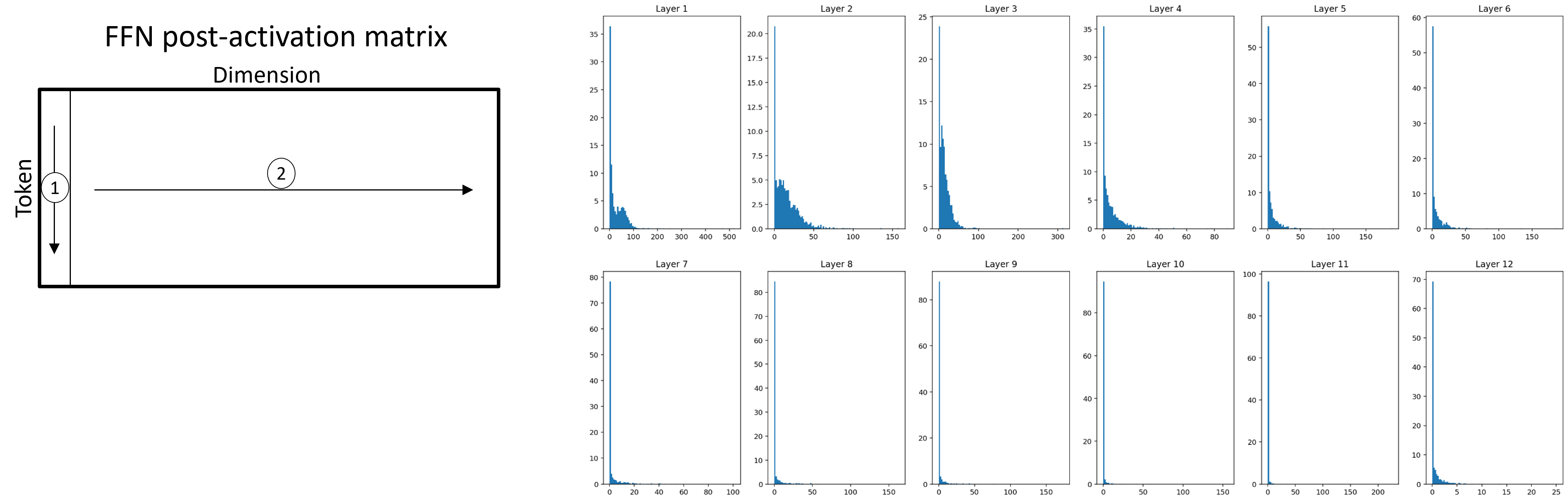}
\caption{Histogram of FFN post-activation accumulation along dimensions}
\label{accumulation}
\end{figure}

Fig.~\ref{accumulation} shows the result distribution when accumulating values per dimension. In the middle and later layers, most dimensions approach zero, with only a few having significantly larger values. This motivates a selective pruning approach targeting dimensions with lower accumulated values.

\subsubsection{Proposed FFN2 Weight Pruning}

\begin{figure}[htbp]
\centering
\includegraphics[width=1.0\linewidth, keepaspectratio=true]{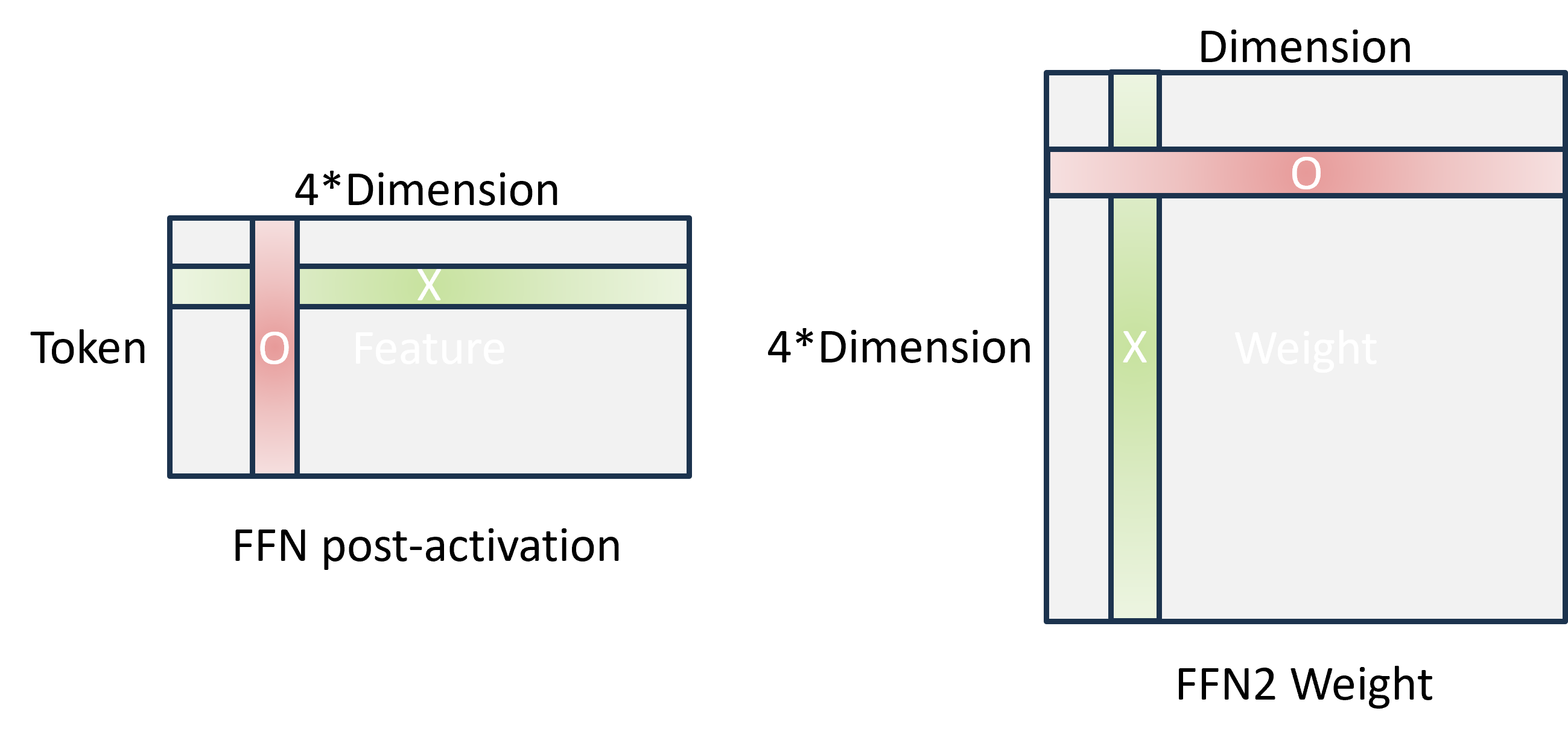}
\caption{Matrix multiplication between FFN activations and FFN2 weights}
\label{matrix_pruning}
\end{figure}

Based on these observations, we propose a dynamic FFN2 weight pruning method that prunes a group of weights with their sum below a threshold. Fig.~\ref{matrix_pruning} illustrates two pruning strategies: group in a token-level or dimension-level. Token-level pruning does not reduce total weight parameters since all weights remain necessary for remaining tokens (green section). Instead, we adopt dimension-level pruning, where pruning an input column allows corresponding weight rows to be pruned (red section), optimizing efficiency.

While FFN1 exhibits similar sparsity, we do not apply pruning to FFN1 due to its direct matrix multiplication structure, which lacks clear preemptive optimization opportunities. Therefore, our focus remains on FFN2, where sparsity-driven weight pruning is more effective.
\subsection{Algorithm-Hardware Co-Design Principles}
The development of our ViT accelerator was guided by an algorithm-hardware co-design philosophy. Algorithmic choices were made with hardware feasibility and efficiency as primary considerations, while the hardware architecture was tailored to support these optimized algorithms effectively. Key aspects include:
\begin{enumerate}
    \item Targeted Optimization: Algorithmic analysis (Fig.~\ref{ViT_Analysis}) identified FFNs as the primary bottleneck for our target ViTs, directing hardware efforts towards FFN-specific optimizations rather than solely focusing on attention mechanisms.
    \item Pruning for Hardware: Both dynamic token pruning (Sec.~\ref{Dynamic Token Pruning}) and FFN2 weight pruning (Sec.~\ref{Dynamic FFN2 Weight Pruning}) were designed to be 'hardware-friendly.' This involved selecting pruning criteria (class attention scores, accumulated dimension sums) and mechanisms (direct discard, simple thresholding) that translate to low-overhead hardware modules (Sec.~\ref{Token Pruning},\ref{FFN2 Pruning}), avoiding complex learnable gates. The choice of dimension-level FFN2 pruning directly facilitates weight memory reduction in hardware.
    \item Activation Function for Simplicity and Sparsity: Replacing GELU with ReLU (Sec.~\ref{Hardware-Friendly Activation}) simplified the activation unit in hardware and, crucially, increased activation sparsity, which is leveraged by our FFN2 pruning algorithm.
    \item Dataflow Co-optimization: The proposed row-wise, output-oriented dataflow and interleaved FFN computation (Sec. ~\ref{Dataflow and detailed mapping}) are algorithmic scheduling and data arrangement strategies specifically designed to minimize on-chip memory, reduce data transposition overhead, and improve data reuse within the hardware architecture. This iterative consideration of algorithmic impact on hardware complexity and hardware constraints on algorithmic choices was central to achieving the reported efficiency.
\end{enumerate}

\section{Hardware Design}
\label{chapter:Hardware}

\begin{figure}[htbp]
\centering
\includegraphics[height=!,width=1.0\linewidth,keepaspectratio=true]{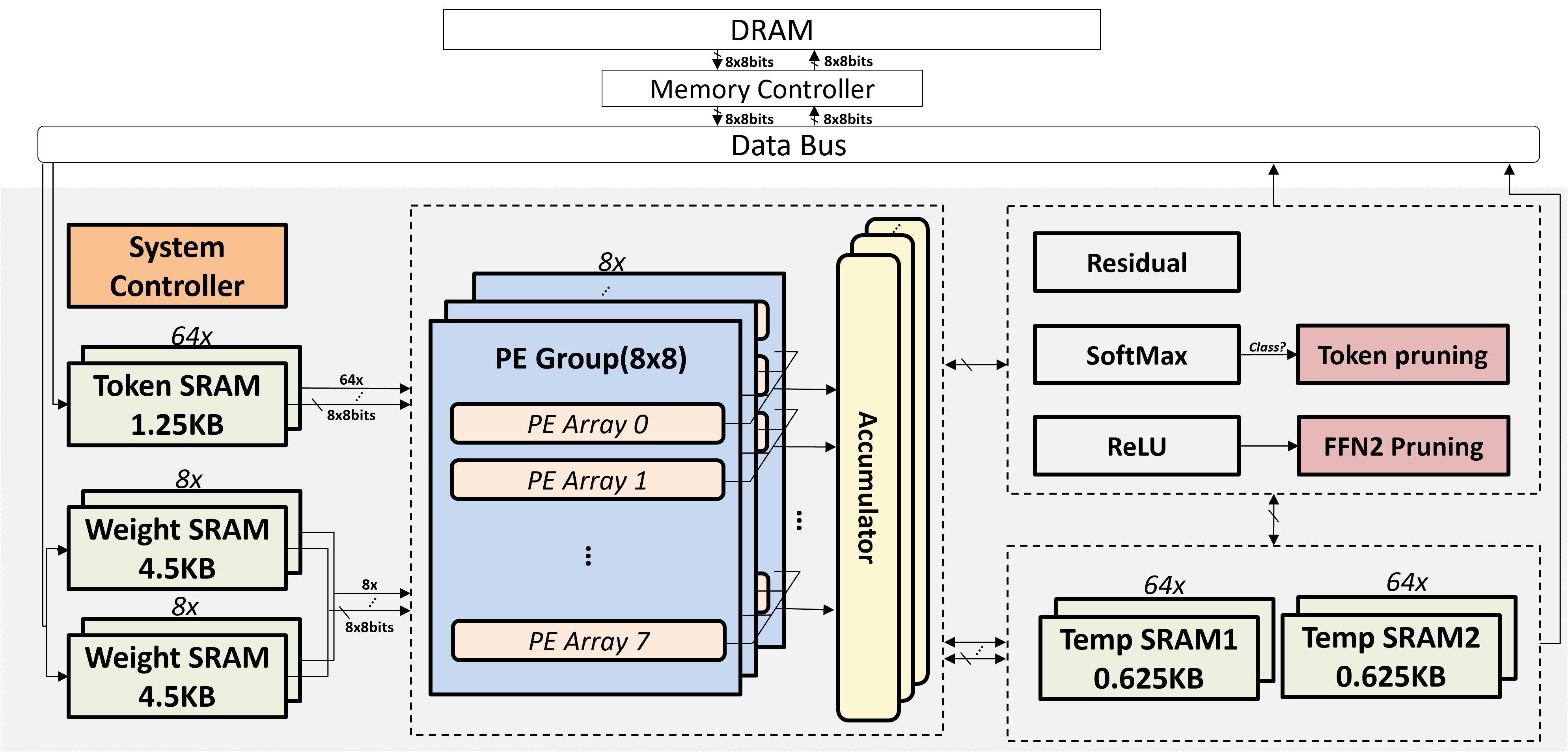}
\caption{The overview of hardware architecture}
\label{Hardware_architecture}
\end{figure}

A key challenge in transformer acceleration is minimizing redundant memory access and ensuring efficient data reuse. Fig.~\ref{Hardware_architecture} shows the proposed hardware accelerator. Our hardware employs a row-wise, output-oriented dataflow strategy that eliminates unnecessary data transpositions, significantly reducing SRAM read/write operations.  In this design, input from token SRAM and weights from weight SRAM are multiplied and accumulated with eight PE groups. Dynamic pruning is applied through Token Pruning and FFN2 Pruning to reduce unnecessary computations, further enhancing efficiency while minimizing any significant loss in accuracy. The whole dataflow is reconfigured for different layer processings.

\subsection{PE}

\begin{figure}[htbp]
\centering
\includegraphics[height=!,width=1.0\linewidth,keepaspectratio=true]{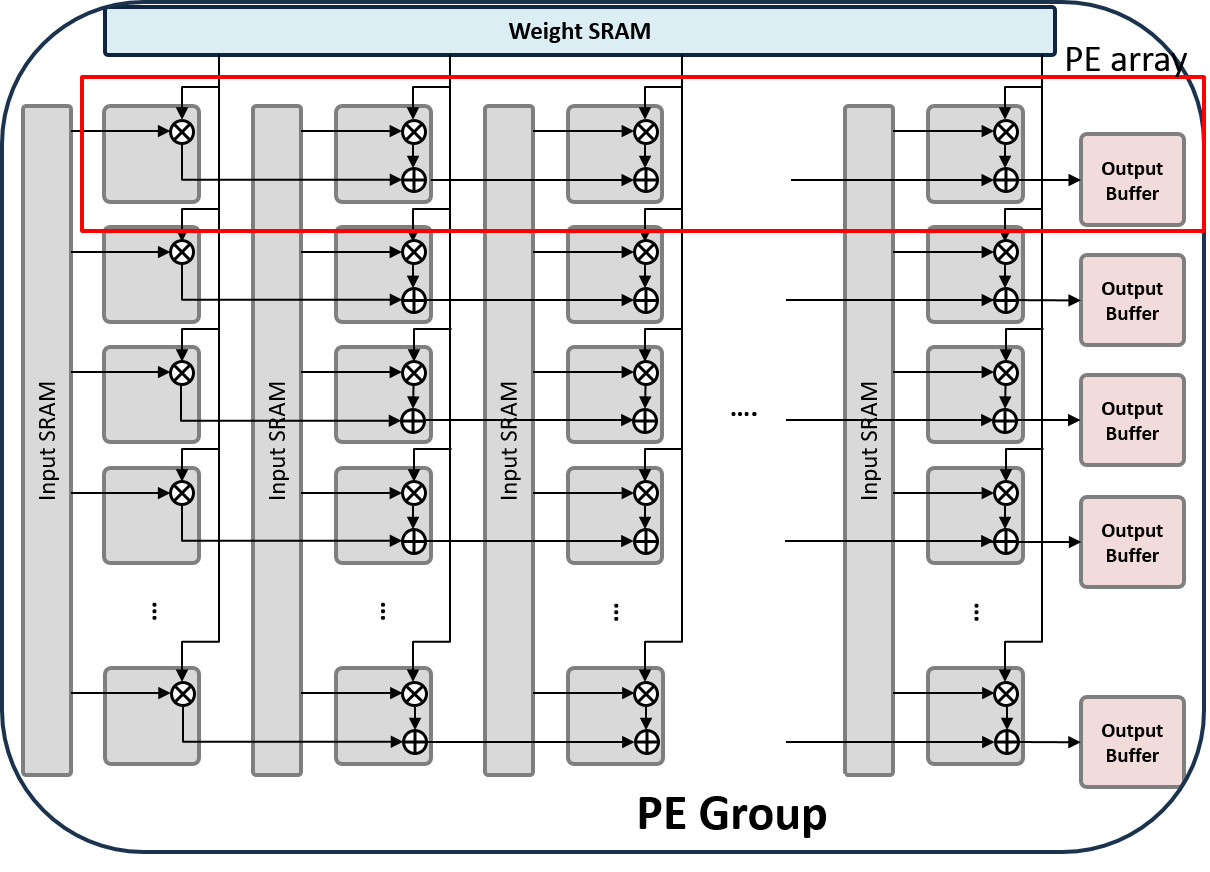}
\caption{Detailed diagram of a PE Group(8x8)}
\label{PE}
\end{figure}

Fig.~\ref{PE} illustrates a PE group consisting of 8 arrays, each containing 8 MAC units. The design employs a column-wise weight broadcasting mechanism to maximize weight reuse, while each MAC unit receives distinct inputs to support matrix multiplication.

The outputs from each PE are accumulated row-wise into a buffer. With 8 PE groups operating concurrently, each group computes a partial result for the same output element at a corresponding row position. These partial results are stored in the buffer and later combined to ensure accurate aggregation of the final output elements. Notably, the architecture is designed to output 8 elements at a time, maximizing parallelism and computational efficiency.

This design choice is motivated by the structure of matrix multiplications in ViT encoder blocks, where intermediate dimensions are multiples of 64. To align with this, we ensure that the product of the accumulation count within a single PE array and the parallelism across the PE groups equals 64. This alignment optimizes computational efficiency.

Additionally, processing 8 elements per cycle offers a significant advantage, despite minor utilization loss when the number of tokens is not a multiple of 8. This approach allows the architecture to achieve a peak throughput of 512 MACs, aligning with the performance of most existing transformer accelerators. This consistency facilitates straightforward benchmarking while ensuring the accelerator meets performance requirements effectively.
\subsection{Dataflow and detailed mapping}
\label{Dataflow and detailed mapping}

We decompose the operations within the Transformer encoder block into two main parts. The first part consists of fully connected operations, including QKV generation, linear projection, and the FFN. The second part focuses on self-attention computation, which is crucial for capturing dependencies across tokens. Both parts involve matrix multiplication and therefore follow the same computational flow, as described in the previous subsection. However, the input access order is arranged in a row-wise or column-wise manner based on the output order, facilitating data access between layers without requiring explicit data transposition. Inspired by~\cite{ViTA}, we implemented a head-level pipeline that processes each head sequentially, ensuring efficient head-by-head computation.

\subsubsection{Fully connected operations}

\begin{figure}[htbp]
\centering
\includegraphics[height=!,width=0.6\linewidth,keepaspectratio=true]{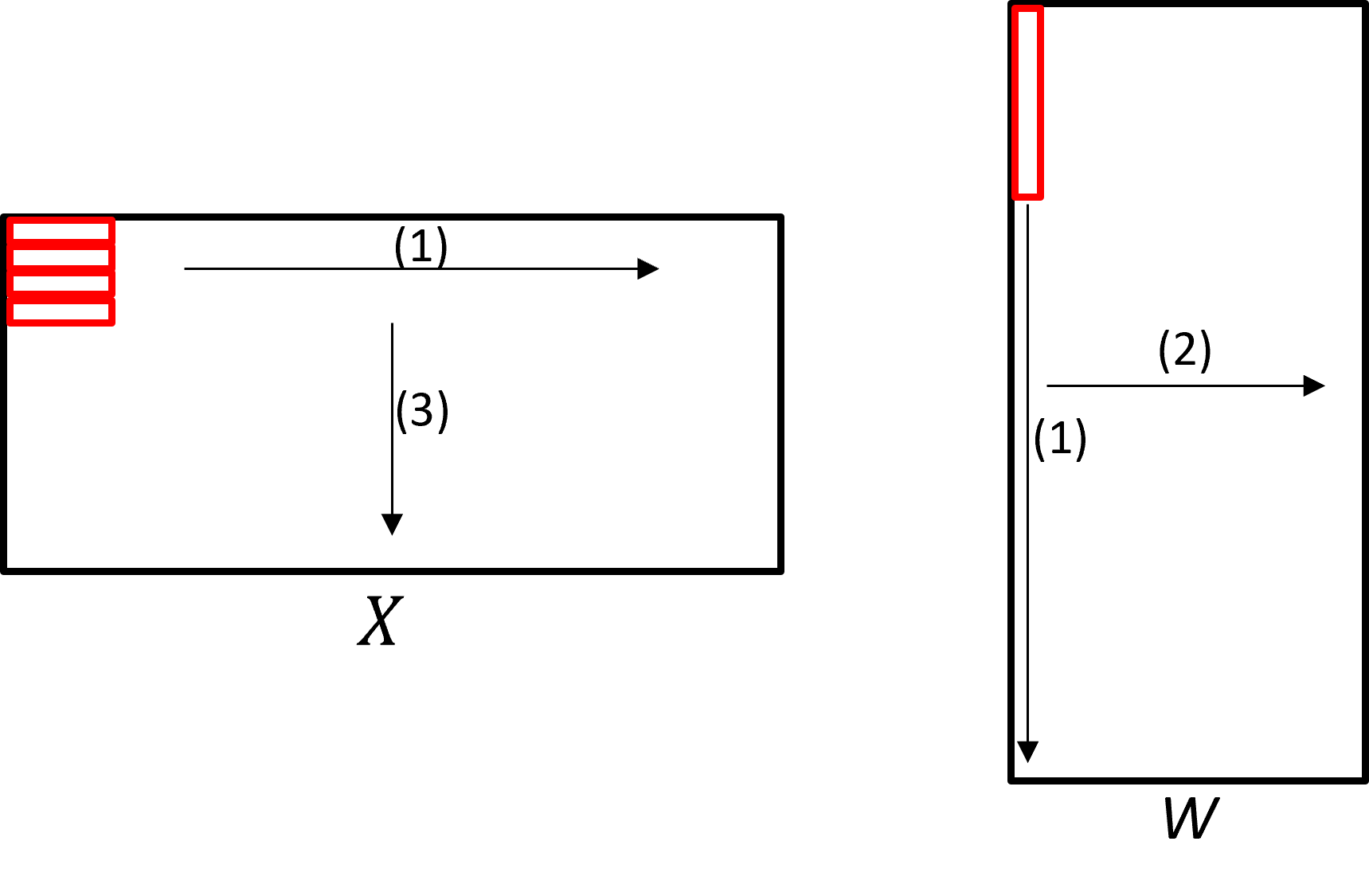}
\caption{Data access flow of fully connected operations.}
\label{Fully_Connect}
\end{figure}

\begin{figure}[htbp]
        \centering
        \includegraphics[height=!,width=1.0\linewidth,keepaspectratio=true]{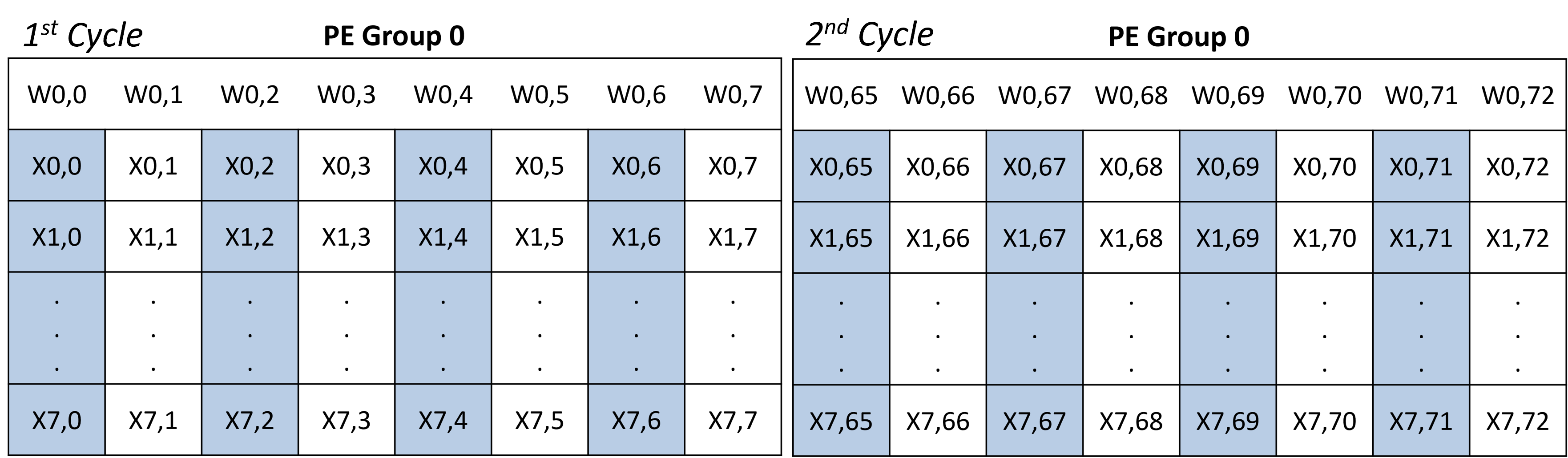}
    \caption{Detailed mapping of the fully connected operations for a PE group at the first and second cycle. Each grid corresponds to one MAC.}
    \label{PE_Group0_FC}
\end{figure}

The dataflow for the fully connected operations is illustrated in Fig.~\ref{Fully_Connect}, covering all matrix multiplications except self-attention. Its data access order depends on the output order. For row-wise outputs as used in Q generation and linear projection, the input \(X\) is processed row-wise, while the weights \(W\) are processed column-wise in the multiplication of \(X\) and \(W\), as denoted by (1) in Fig.~\ref{Fully_Connect}. For column-wise output as used in KV generation and FFN1, their order will be interchanged as denoted by (2) and (3) in Fig.~\ref{Fully_Connect}. The output order is selected to ease data access between different layers.

Fig.~\ref{PE_Group0_FC} depicts the mapping of inputs and weights, which are updated in subsequent cycles. This design processes 64 weights per cycle. If the total number of weights exceeds 64, additional accumulation cycles are required. For instance, when processing 384 weights, 6 accumulation cycles are needed. After these 6 cycles, 8 elements are output simultaneously.

\subsubsection{FFN}

\begin{figure}[htbp]
\centering
\includegraphics[height=!,width=0.6\linewidth,keepaspectratio=true]{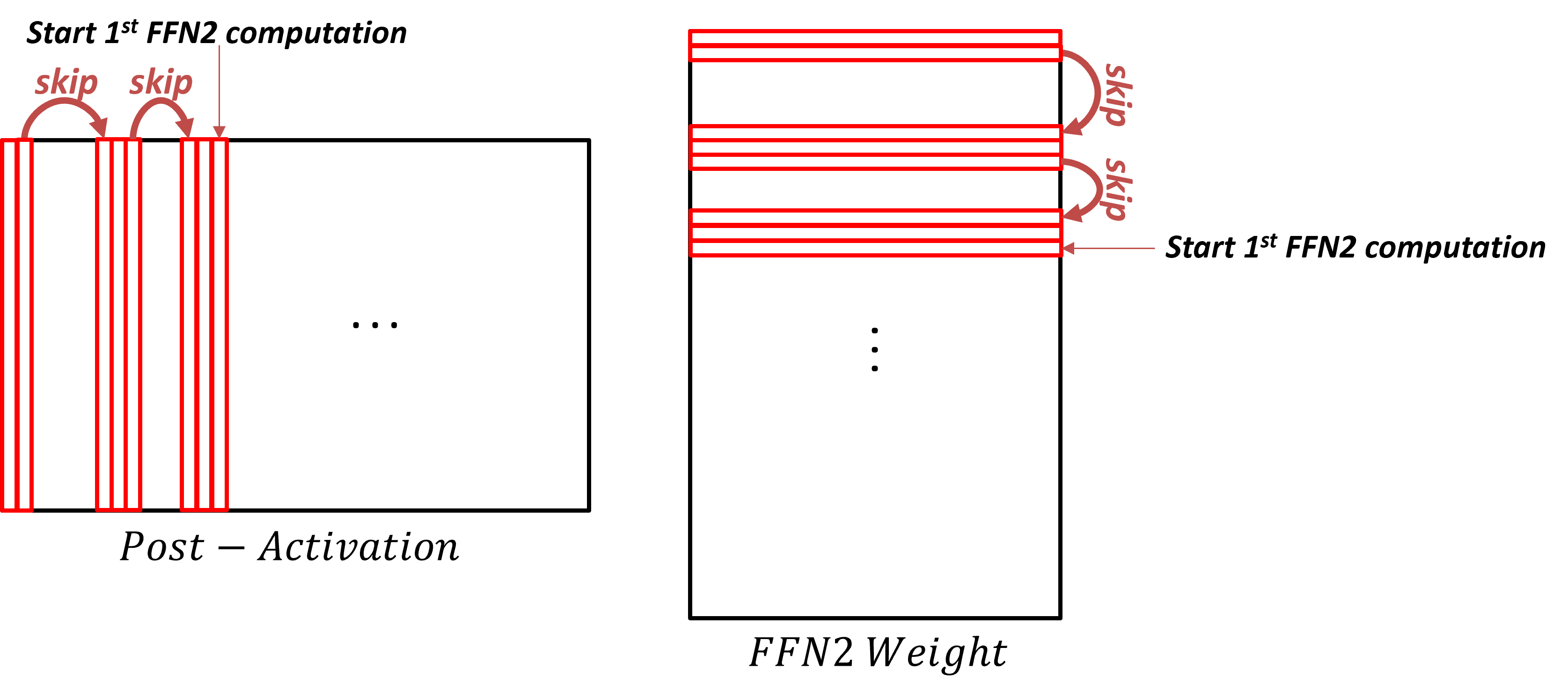}
\caption{Data access flow of FFN2}
\label{FFN}
\end{figure}

In the FFN computation, the FFN1 output need to be stored for the following FFN2. To avoid a larger intermediate buffer, we propose an interleaved FFN computing order. In this order, once FFN1 computations are completed, the results are passed through the ReLU activation function before entering the FFN2 pruning module, which determines whether the computed elements should be stored in the Temp SRAM. Next, a predefined number of FFN2 weights are fetched to perform FFN2 computations.
As illustrated in Fig.~\ref{FFN}, after computing the eight columns of the post-activation outputs, the corresponding eight rows of FFN2 weights are loaded to initiate FFN2 computation. The pruned weight will be skipped. Once a partial FFN2 computation is completed, the process resumes with the remaining FFN1 computations. This alternating pattern continues, where FFN1 and FFN2 computations are interleaved, allowing partial sums to be accumulated efficiently.

This interleaved approach ensures efficient data processing by carefully managing intermediate results while optimizing both computational performance and memory usage. The outputs from these computations are temporarily stored in Temp SRAM until the entire processing cycle is completed.

\subsubsection{Self-attention}

\begin{figure}[htbp]
\centering
\includegraphics[height=!,width=0.6\linewidth,keepaspectratio=true]{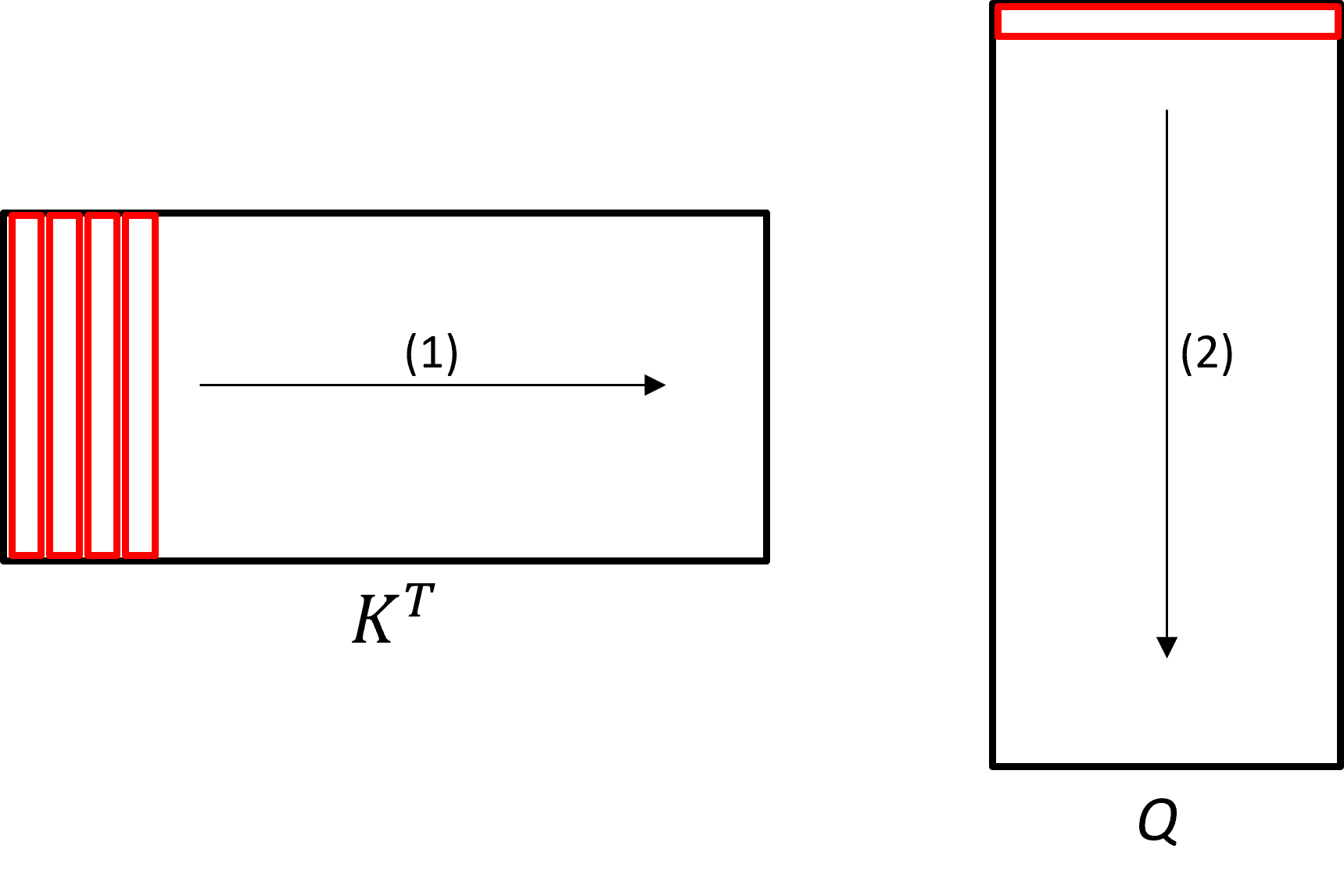}
\caption{Data access flow of \begin{math}Q \times K^T\end{math}}
\label{QK}
\end{figure}

\begin{figure}[htbp]
        \centering
        \includegraphics[height=!,width=1.0\linewidth,keepaspectratio=true]{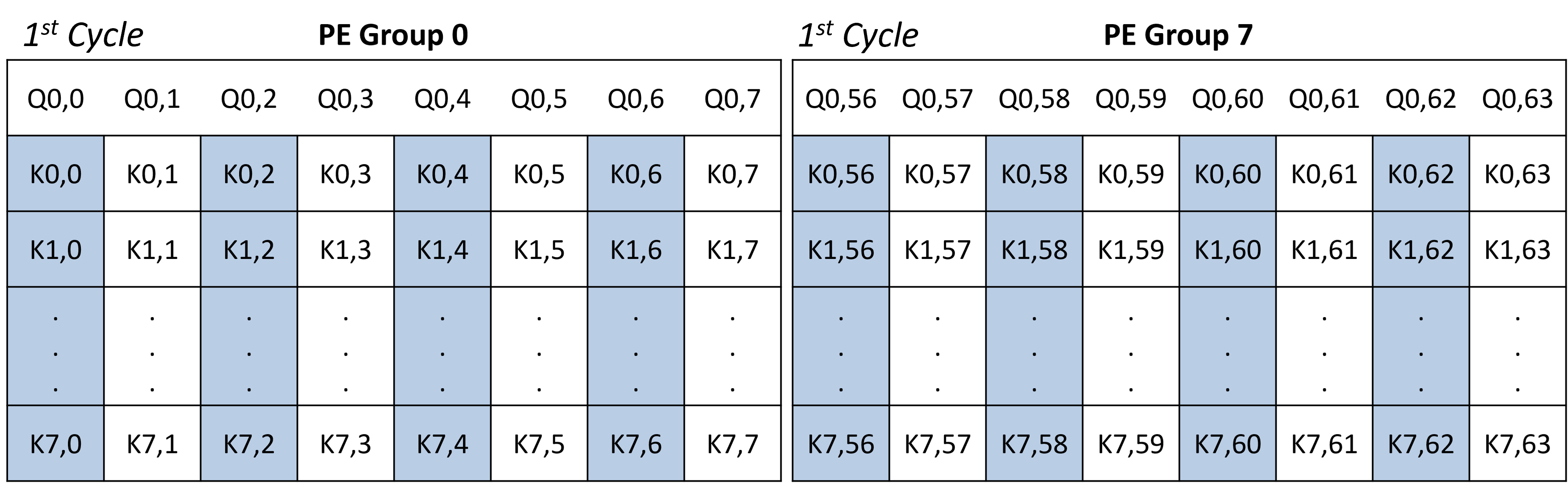}
    \caption{Detailed mapping of \begin{math}Q \times K^T\end{math}  of (a) PE Group 0 and (b) PE Group 7 at the same cycle.}
    \label{PE_Group0_QK}
\end{figure}

Fig.~\ref{QK} illustrates the data access flow of self-attention for computing the \( Q \times K^T \) matrix. The shared dimension between \( Q \) and \( K \) is exactly 64, enabling the weights to be mapped across all PE Groups in a single cycle. This efficient mapping, as shown in Fig.~\ref{PE_Group0_QK}, allows the computation to output 8 elements in just one cycle. In this flow, we set Q as weight that also uses the column-wise weight broadcasting mechanism. The output is row-wise to align with the SoftMax function's requirements, and thus the data access order of Q and K follows the directions as in Fig.~\ref{QK}.

After the softmax operation, the attention scores are multiplied by \( V \), following a process similar to that of the fully connected layer. However, accumulating all partial sums for a single element in one cycle is not feasible. Therefore, we employ a row-wise dataflow here, similar to the one used for \( Q \times K^T \). This method ensures efficient computation and data handling, optimizing the utilization of hardware resources throughout the process.

\subsection{Token Pruning}
\label{Token Pruning}
\begin{figure}[htbp]
\centering
\includegraphics[height=!,width=1.0\linewidth,keepaspectratio=true]{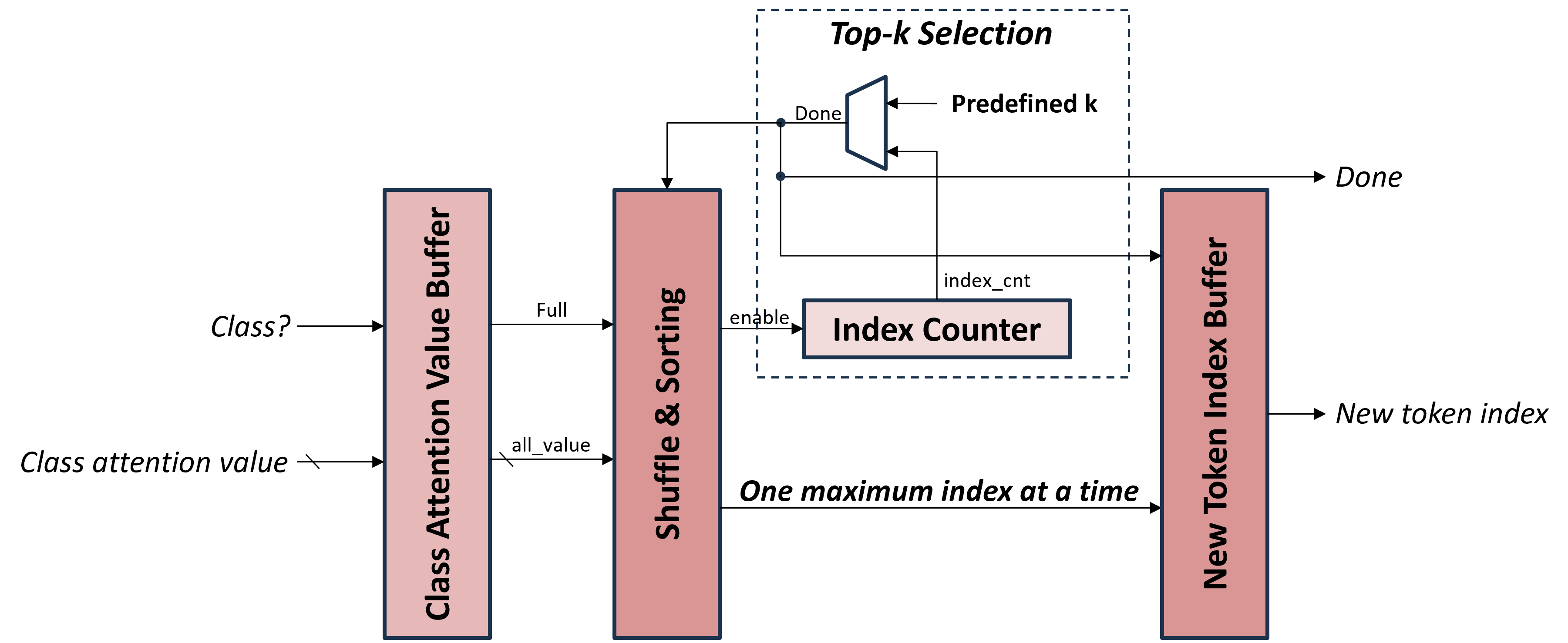}
\caption{Block diagram of token pruning module}
\label{Token_pruning_module}
\end{figure}

Fig.~\ref{Token_pruning_module} illustrates the design of the token pruning module. This module begins by storing the class attention values in a buffer. Once all values are available, they are passed to the sorting block for processing.

The sorting block, inspired by hardware-efficient sorting methodologies~\cite{sorting, parallell-sorting}—which leverage bit-level manipulation to reduce comparator complexity and enable comparison-free stages—was selected for its suitability for pipelined hardware implementation. The input data is initially shuffled to group corresponding bits from each data point together. This shuffling is essential to ensure the correct operation of the sorting algorithm. Once shuffled, the sorting block processes the data and outputs, in each cycle, the index corresponding to the maximum value. This index is then forwarded to the new token buffer, where it is reordered for subsequent operations.

To further optimize the pruning process in line with our algorithmic requirements, we have integrated a Top-$K$ selection mechanism into the token pruning module. The sorting block operates until it outputs $K$ indices, where $K$ is computed as $K = \lceil (N-1) \times \rho \rceil$, with $N$ representing the total number of tokens and $\rho$ denoting the keep ratio. This calculation determines the number of tokens to retain, excluding the class token. This approach enables the identification of the top-$K$ token indices with a deterministic latency proportional to $K$ after an initial setup phase, fitting within the timing constraints of layer-wise processing in ViTs such as DeiT-S ($N=197$ tokens, $K \approx 98$). While fully parallel sorters (e.g., bitonic sorters) can achieve $\mathcal{O}(\log^2 N)$ latency, they incur an $\mathcal{O}(N \log^2 N)$ hardware cost. In contrast, our chosen method offers a favorable trade-off by sequentially identifying the $K$ most important indices, which is efficient since only these $K$ tokens are propagated for further computation.

\subsection{FFN2 Pruning}
\label{FFN2 Pruning}
\begin{figure}[htbp]
\centering
\includegraphics[height=!,width=1.0\linewidth,keepaspectratio=true]{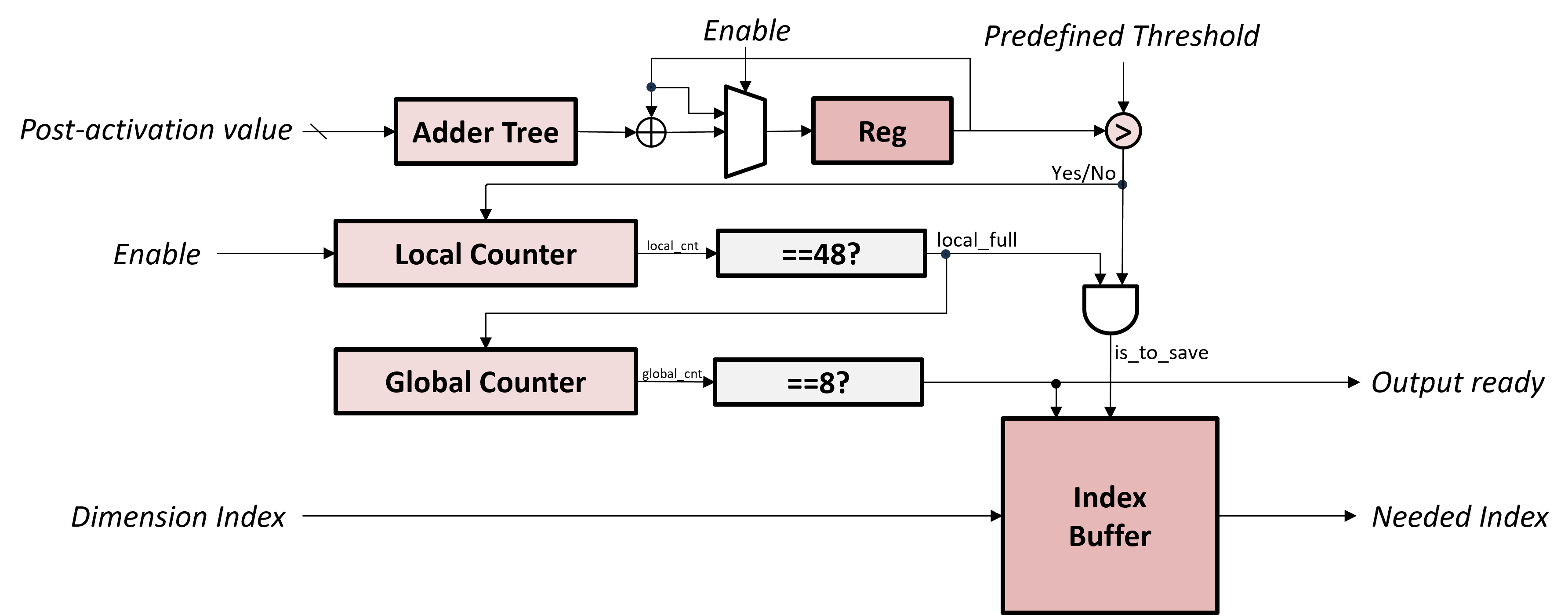}
\caption{Block diagram of the FFN2 pruning module}
\label{FFN2_pruning}
\end{figure}

The FFN2 pruning module is designed to be even simpler than the token pruning module, as it does not require any sorting operations. Instead, this module operates by comparing the accumulated value for each dimension against a predefined threshold. If the value exceeds the threshold, the corresponding index is retained; otherwise, it is discarded.

The module utilizes a Local Index Counter to track whether a complete dimension has been accumulated. As shown in Fig.~\ref{FFN2_pruning}, 8 full elements are processed at a time, meaning the module must accumulate over 48 times to cover the entire dimension in FFN, as in DeiT-S, for example. Once the local counter confirms that all necessary values for one dimension have been processed, the Global Index Counter checks whether 8 indices that need to be retained have been identified. Upon reaching this count, the module starts outputting the indices from the Index Buffer to fetch weights from external memory.

This design streamlines the pruning process, making it highly efficient by focusing on simple comparisons rather than complex operations while still ensuring that only the most significant dimensions are retained for further processing.

\section{Experimental Result}
\label{chapter:Experimental Result}

\subsection{Experimental Setup}
The proposed model modifications were integrated into DeiT~\cite{DeiT} and trained on the ImageNet-1k~\cite{imagenet} dataset for image classification using four A100 GPUs. 
The training process employed the AdamW optimizer over 300 epochs, with a batch size of 512 per GPU, resulting in a total batch size of 2048 across all GPUs. The initial learning rate was set to \(2 \times 10^{-3}\) and adjusted using cosine learning rate scheduling. Cross-entropy was used as the loss function, with a token pruning rate of 0.5.
For fine-tuning, the warmup phase was removed, the initial learning rate was set to \(2 \times 10^{-5}\), and weight decay was adjusted to \(1 \times 10^{-6}\). The number of epochs was modified as necessary, while other configurations remained unchanged.

The proposed hardware design was implemented in SystemVerilog and synthesized using Synopsys Design Compiler, targeting a 28nm CMOS technology. The design operates at 1 GHz, with power consumption evaluated using Synopsys PrimeTime PX.

\subsection{Analysis of Model Optimization}

\begin{figure}[htbp]
    \centering
    \includegraphics[width=0.7\linewidth,keepaspectratio]{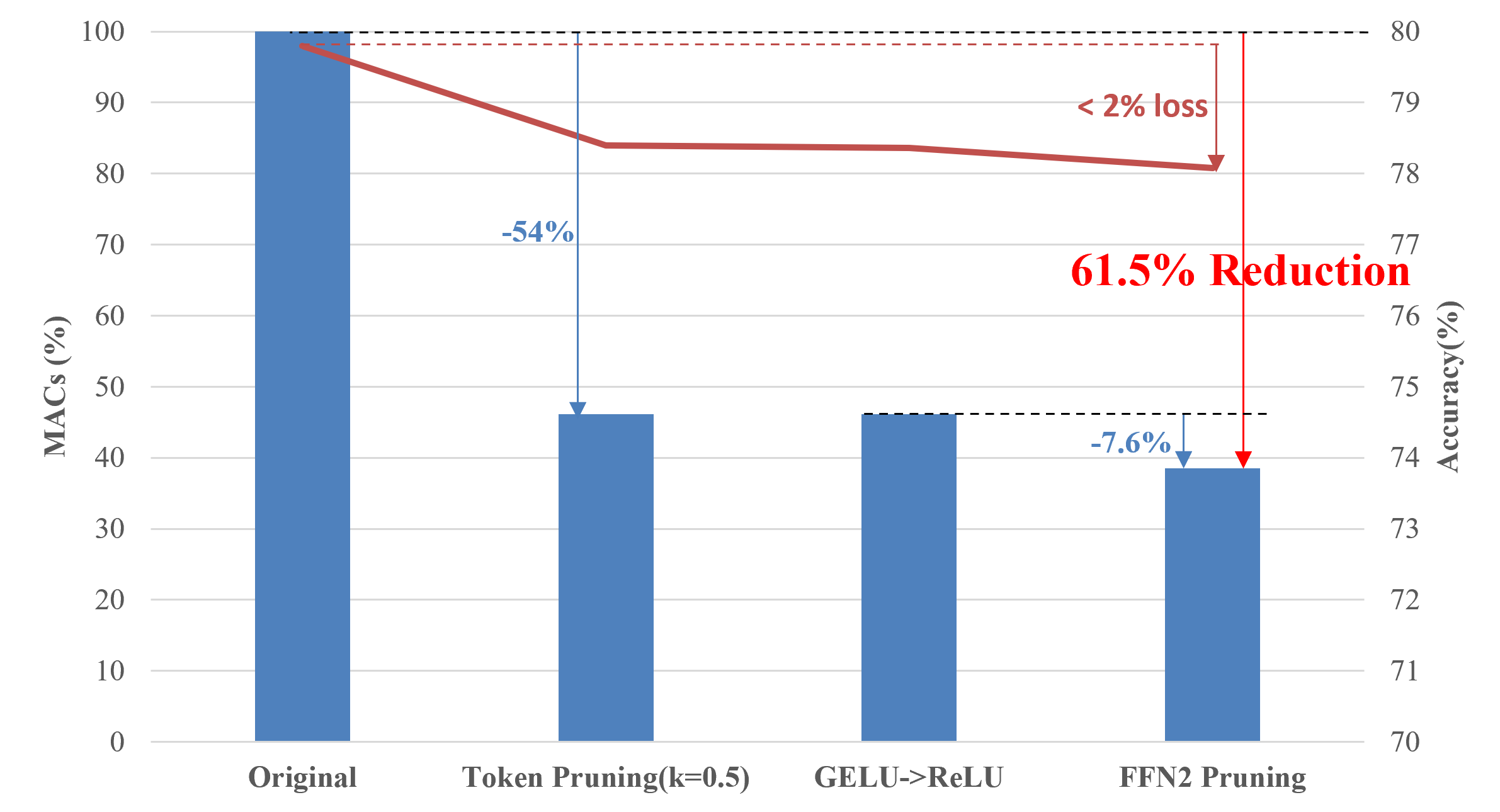}
    \caption{Analysis of MAC reduction.}
    \label{performance}
\end{figure}

Fig.~\ref{performance} presents the MAC reduction analysis for the proposed techniques. Dynamic token pruning alone, with a keep ratio of $\rho$ = 0.5 for Top-$K$ pruning, reduced MACs by 54\%, while substituting GELU with ReLU and applying FFN2 weight pruning further reduced computations by 7.6\%. The combined optimizations led to a total reduction of 61.5\% while maintaining 78.08\% accuracy.

\begin{figure}[htbp]
    \centering
    \includegraphics[width=0.6\linewidth,keepaspectratio]{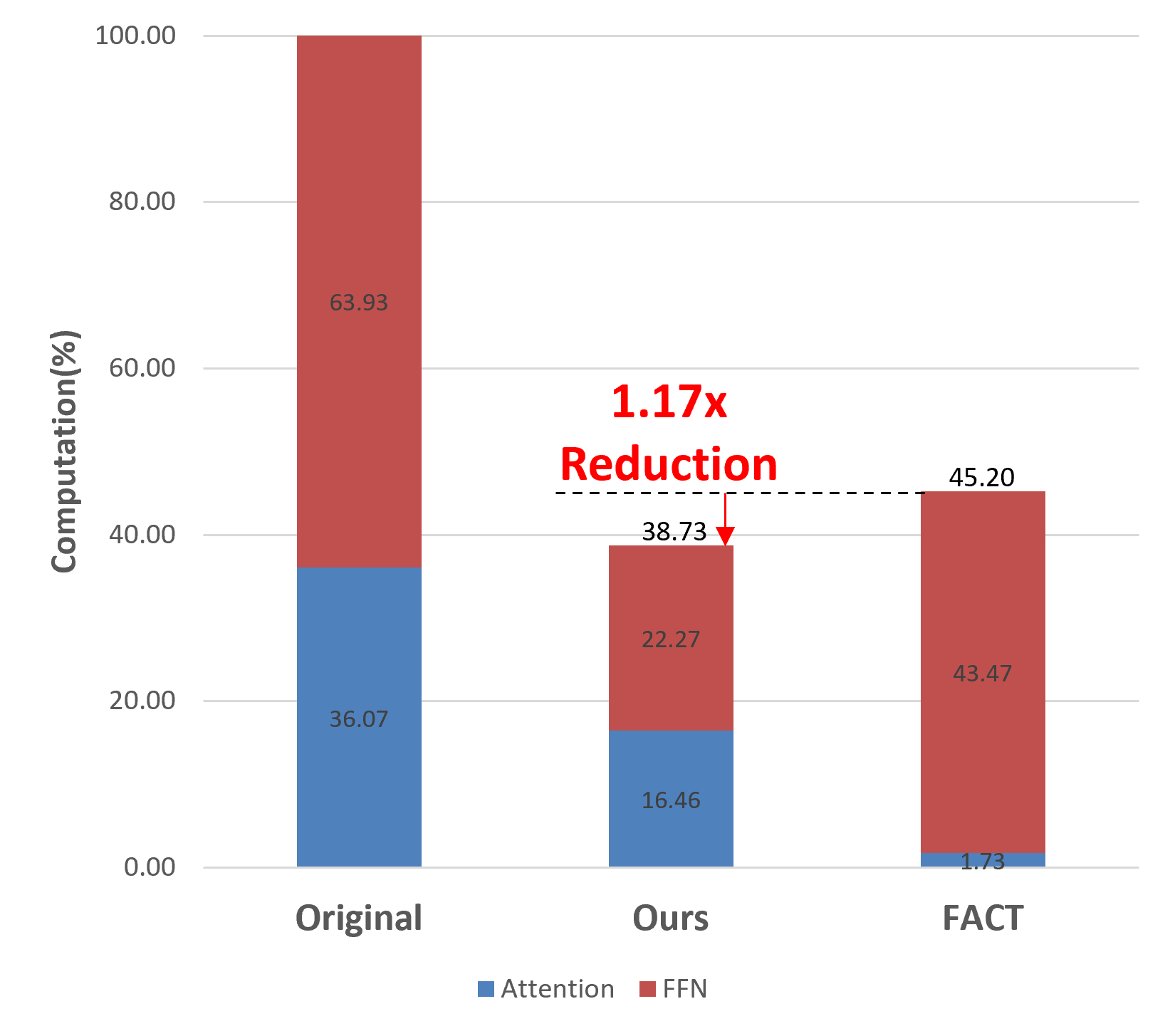}
    \caption{Computation comparison with FACT~\cite{FACT}.}
    \label{performance_comp}
\end{figure}

Fig.~\ref{performance_comp} compares our approach to the FACT accelerator~\cite{FACT} for the ViT-B/16 model. Although FACT achieves substantial savings in the attention mechanism, our approach improves overall computational reduction by 1.17x while maintaining accuracy loss within the 2\% threshold.

\subsection{Hardware Implementation Result}

\subsubsection{Memory Access Reduction for Tokens}

To analyze external memory access, we assess the impact of token pruning on the amount of data fetched during computation. Each layer requires fetching tokens at least once for self-attention and FFN computations. Aggregating this across all 12 layers in DeiT-S allows us to measure total token fetch requirements.

\begin{figure}[htbp]
\centering
\includegraphics[height=!,width=0.8\linewidth,keepaspectratio=true]{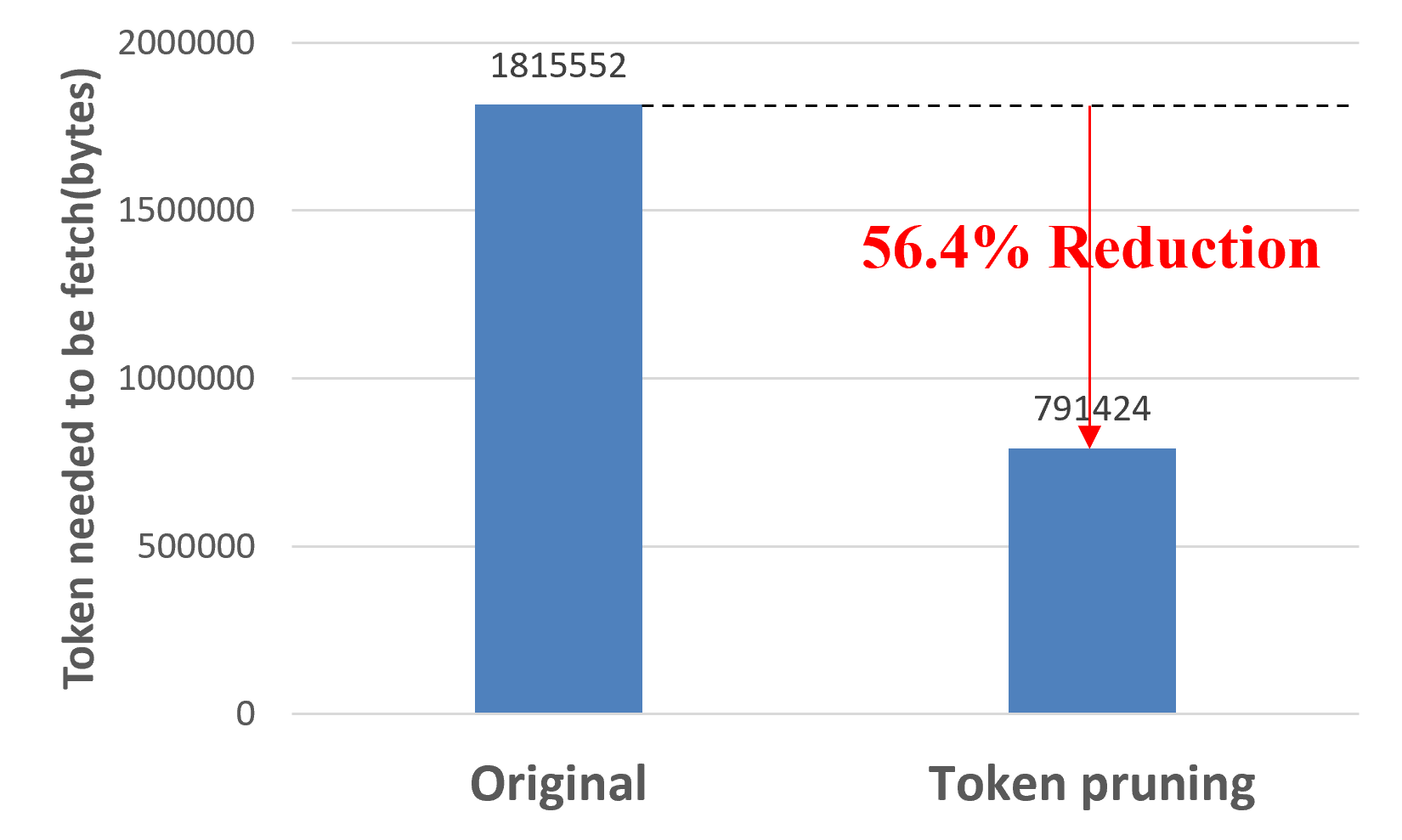}
\caption{External memory access for token}
\label{fetch}
\end{figure}

\begin{figure}[htbp]
\centering
\includegraphics[height=!,width=0.8\linewidth,keepaspectratio=true]{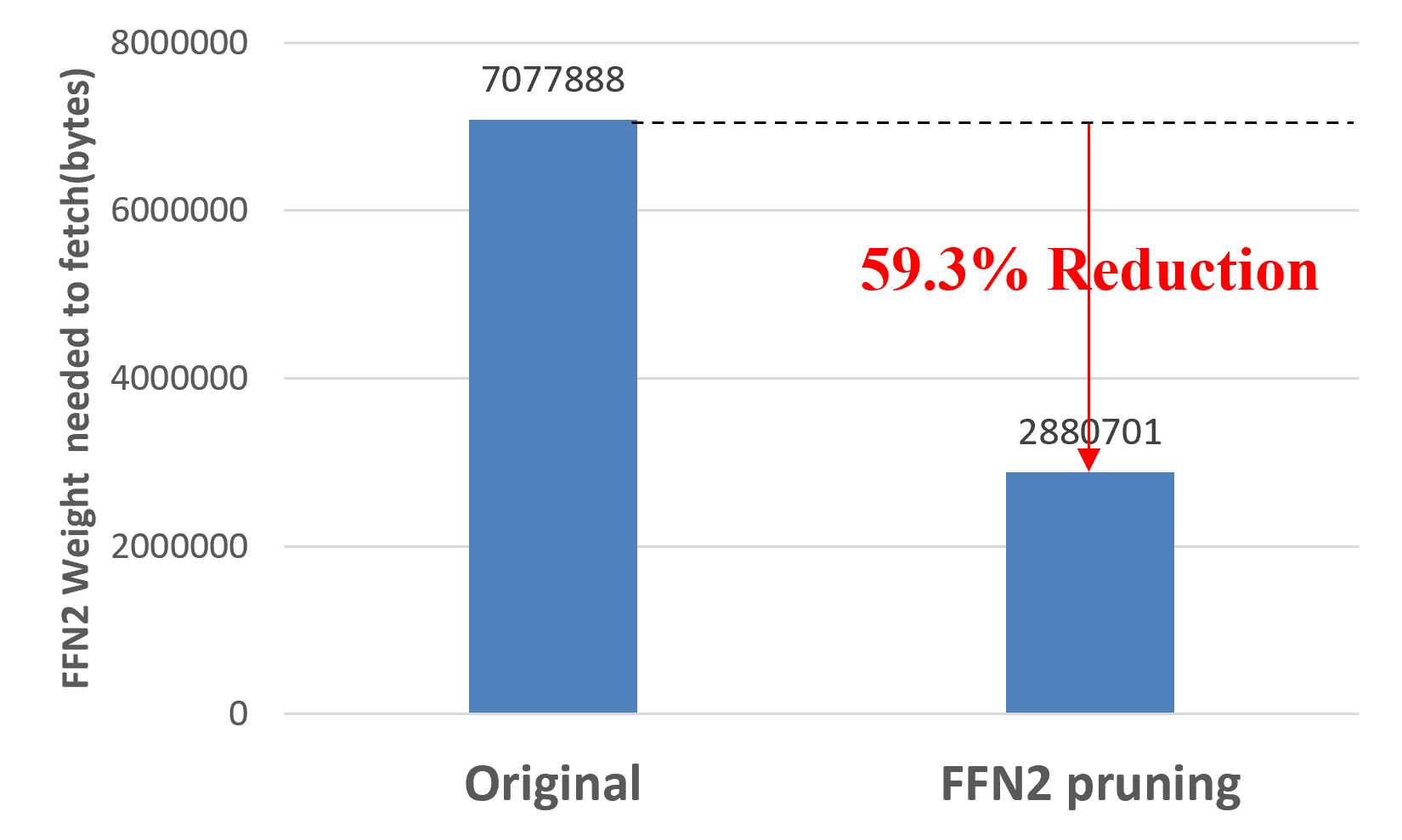}
\caption{External memory access for FFN2 weight}
\label{fetch_FFN2}
\end{figure}

\begin{figure}[htbp]
\centering
\includegraphics[height=!,width=0.8\linewidth,keepaspectratio=true]{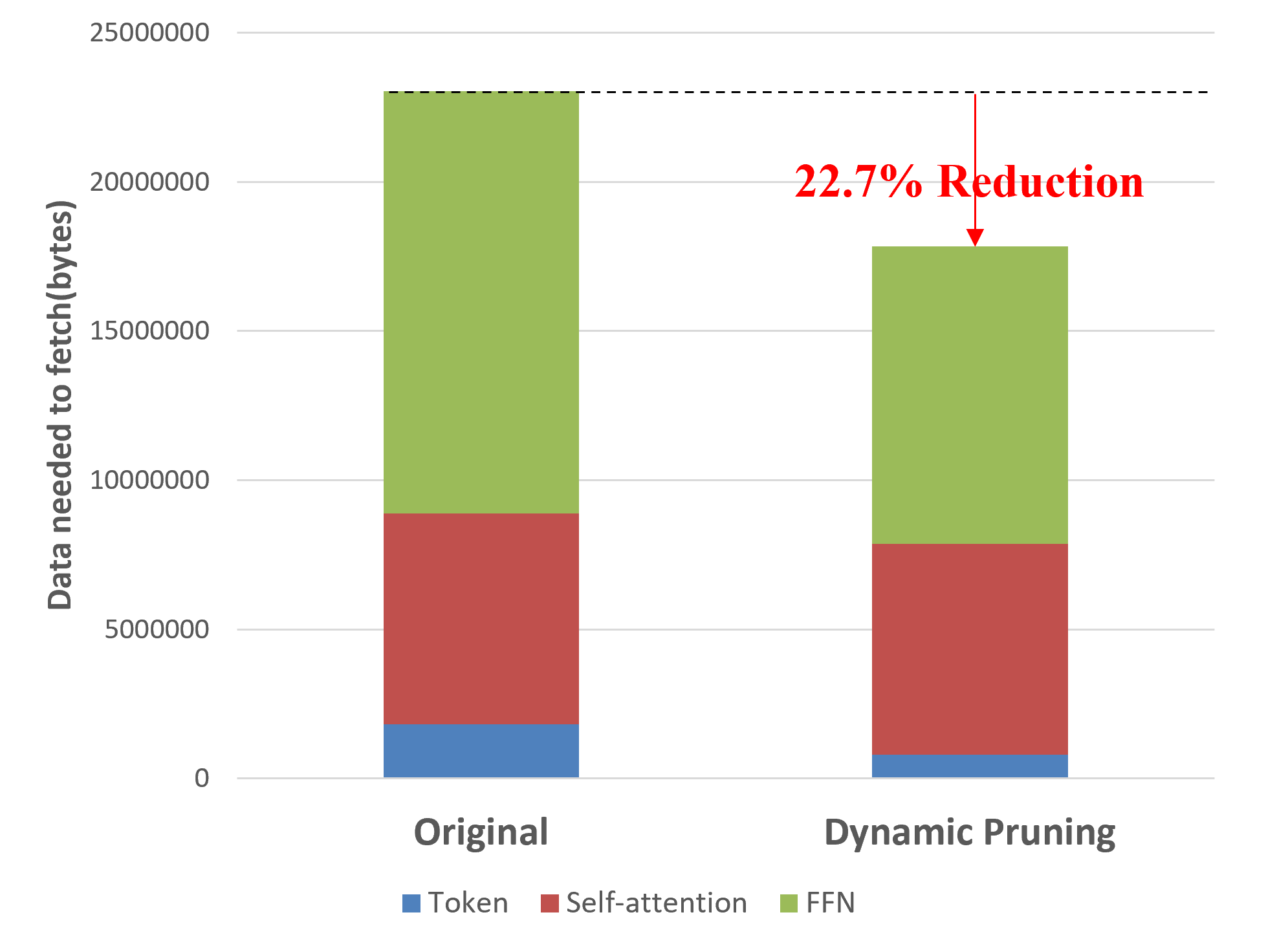}
\caption{External memory access with dynamic Pruning}
\label{fetch_total}
\end{figure}

Fig.~\ref{fetch} illustrates a 56.4\% reduction in total data fetch requirements when applying token pruning with \(\rho = 0.5\). This significantly reduces computational overhead while improving hardware efficiency. In addition, Fig.~\ref{fetch_FFN2} shows the reduction in FFN2 weight fetches by 59.3\% due to our layer-adaptive thresholding strategy. By combining token and FFN2 pruning, we achieve a total 22.7\% reduction in overall memory fetch requirements, as shown in Fig.~\ref{fetch_total}.

\begin{table*}[htbp]
\centering
\caption{Comparison with other designs.}
\label{design_comparison}
\begin{tabular}{|ccccccc|}
\hline
\multicolumn{1}{|c|}{}                                                                                                         & \multicolumn{1}{c|}{This work}      & \multicolumn{1}{c|}{\cite{FACT}}          & \multicolumn{1}{c|}{\cite{28nm}}                        & \multicolumn{1}{c|}{\cite{SpAtten}}                & \multicolumn{1}{c|}{\cite{A3}}              & \cite{ELSA}                   \\ \hline
\multicolumn{1}{|c|}{Technology(nm)}                                                                                               & \multicolumn{1}{c|}{28}             & \multicolumn{1}{c|}{28}            & \multicolumn{1}{c|}{28}                          & \multicolumn{1}{c|}{40}                     & \multicolumn{1}{c|}{40}                                & 40                     \\ \hline
\multicolumn{1}{|c|}{Status} & \multicolumn{1}{c|}{Simulation} & \multicolumn{1}{c|}{Chip} & \multicolumn{1}{c|}{Chip} & \multicolumn{1}{c|}{Simulation} & \multicolumn{1}{c|}{chip} & \multicolumn{1}{c|}{Simulation} \\ \hline
\multicolumn{1}{|c|}{Application}                                                                                              & \multicolumn{1}{c|}{Vision}         & \multicolumn{1}{c|}{Vision+NLP}    & \multicolumn{1}{c|}{Vision}                      & \multicolumn{1}{c|}{NLP}                    & \multicolumn{1}{c|}{NLP}                               & NLP                    \\ \hline
\multicolumn{1}{|c|}{Optimization}                                                                                             & \multicolumn{1}{c|}{Attention+FFN}  & \multicolumn{1}{c|}{Attention+FFN} & \multicolumn{1}{c|}{Attention}                   & \multicolumn{1}{c|}{Attention}              & \multicolumn{1}{c|}{Attention}                         & Attention              \\ \hline
\multicolumn{1}{|c|}{Precision}                                                                                                & \multicolumn{1}{c|}{INT8}           & \multicolumn{1}{c|}{INT8}          & \multicolumn{1}{c|}{INT12}                       & \multicolumn{1}{c|}{INT12}                  & \multicolumn{1}{c|}{INT9}                              & INT8/FP16              \\ \hline
\multicolumn{1}{|c|}{PE Numbers}                                                                                               & \multicolumn{1}{c|}{512}            & \multicolumn{1}{c|}{512}           & \multicolumn{1}{c|}{512}                         & \multicolumn{1}{c|}{1024}                   & \multicolumn{1}{c|}{512}                               & -                      \\ \hline
\multicolumn{1}{|c|}{Area (KGE)}                                                                                               & \multicolumn{1}{c|}{496.5}          & \multicolumn{1}{c|}{-}             & \multicolumn{1}{c|}{-}                           & \multicolumn{1}{c|}{-}                      & \multicolumn{1}{c|}{680.5}                             & -                      \\ \hline
\multicolumn{1}{|c|}{Area (mm{$^2$})}                                                                             & \multicolumn{1}{c|}{\textbf{1.193}}          & \multicolumn{1}{c|}{6.03}          & \multicolumn{1}{c|}{6.82}                        & \multicolumn{1}{c|}{1.55}                   & \multicolumn{1}{c|}{2.082}                             & 2.15                   \\ \hline
\multicolumn{1}{|c|}{SRAM (KB)}                                                                                                & \multicolumn{1}{c|}{232}            & \multicolumn{1}{c|}{512}           & \multicolumn{1}{c|}{336}                         & \multicolumn{1}{c|}{392}                    & \multicolumn{1}{c|}{80}                                & 148.5                  \\ \hline
\multicolumn{1}{|c|}{Clock Rate (MHz)}                                                                                         & \multicolumn{1}{c|}{1000}           & \multicolumn{1}{c|}{500}           & \multicolumn{1}{c|}{50-510}                      & \multicolumn{1}{c|}{1000}                   & \multicolumn{1}{c|}{1000}                              & 1000                   \\ \hline
\multicolumn{1}{|c|}{{$^d$}Peak Throughput (GOPS)}                                                                                   & \multicolumn{1}{c|}{1024 / {$^a$}2659}           & \multicolumn{1}{c|}{512 / {$^a$}928}     & \multicolumn{1}{c|}{522}                         & \multicolumn{1}{c|}{360}                    & \multicolumn{1}{c|}{220}                               & 1090                   \\ \hline
\multicolumn{1}{|c|}{Power (mW)}                                                                                               & \multicolumn{1}{c|}{443.2}          & \multicolumn{1}{c|}{337.07}        & \multicolumn{1}{c|}{12.06-272.8}                 & \multicolumn{1}{c|}{942.41}                 & \multicolumn{1}{c|}{110.422}                           & 1494.22                \\ \hline
\multicolumn{1}{|c|}{\multirow{2}{*}{\begin{tabular}[c]{@{}c@{}}{$^d$}Energy Efficiency\\ (TOPS/W)\end{tabular}}}                    & \multicolumn{1}{c|}{2.31 / {$^a$}\textbf{6.00}}           & \multicolumn{1}{c|}{1.52 / {$^a$}2.75}   & \multicolumn{1}{c|}{1.91}                        & \multicolumn{1}{c|}{0.38}                   & \multicolumn{1}{c|}{1.99}                              & 0.73                   \\ \cline{2-7} 
\multicolumn{1}{|c|}{}                                                                                                         & \multicolumn{1}{c|}{2.31 / {$^a$}6.00}           & \multicolumn{1}{c|}{{{$^b$}2.27 / $^{ab}$}4.11}   & \multicolumn{1}{c|}{{$^b$}2.85}               & \multicolumn{1}{c|}{{$^b$}1.15}                   & \multicolumn{1}{c|}{{$^b$}\textbf{6.07}}                     & {$^b$}2.23                   \\ \hline
\multicolumn{1}{|c|}{\multirow{2}{*}{\begin{tabular}[c]{@{}c@{}}{$^d$}Area Efficiency\\ (GOPS/mm{$^2$})\end{tabular}}}  & \multicolumn{1}{c|}{\textbf{858.61}}         & \multicolumn{1}{c|}{153.90}        & \multicolumn{1}{c|}{76.54}                       & \multicolumn{1}{c|}{238}                    & \multicolumn{1}{c|}{105.67}                            & 506.98                 \\ \cline{2-7} 
\multicolumn{1}{|c|}{}                                                                                                         & \multicolumn{1}{c|}{858.61}         & \multicolumn{1}{c|}{153.90}        & \multicolumn{1}{c|}{76.54}                       & \multicolumn{1}{c|}{{$^c$}485.71}                    & \multicolumn{1}{c|}{{$^c$}215.65}                            & {$^c$}\textbf{1034.65}                 \\ \hline
\multicolumn{1}{|c|}{Supply Voltage (V)}                                                                                       & \multicolumn{1}{c|}{0.9}            & \multicolumn{1}{c|}{1.1}           & \multicolumn{1}{c|}{0.56-1.1}                    & \multicolumn{1}{c|}{1.1}                    & \multicolumn{1}{c|}{1.1}                               & 1.1                    \\ \hline
\multicolumn{4}{|l}{\multirow{3}{*}{\begin{tabular}[c]{@{}l@{}}{$^a$}Considering overall geomean speedup. \\ {$^b$}Normalization energy efficiency = energy eff.{\scriptsize $\times (\dfrac{process}{28\ nm})^2 \times (\dfrac{voltage}{0.9\ V})^2$}. \end{tabular}}} & \multicolumn{3}{l|}{\multirow{3}{*}{\begin{tabular}[c]{@{}l@{}}{$^c$}Normalization area efficiency = area eff.{\scriptsize $\times (\dfrac{process}{28\ nm})^2$}. \\ {$^d$}Dense computation if not specified. \end{tabular}}} \\
\multicolumn{4}{|l}{}                                                                                                                                                                                     & \multicolumn{3}{l|}{}                                                                                                                                                            \\
\multicolumn{4}{|l}{}                                                                                                                                                                                     & \multicolumn{3}{l|}{}                                                                                                                                                            \\ \hline
\end{tabular}
\end{table*}

\begin{table}[htbp]
\centering
\caption{Result of removing token fusion. MACs is calculated with only transformer encoder block excluding Layernorm.}
\begin{tabular}{ccc}
\hline
\textbf{Method}                   & \textbf{ImageNet-1k Top-1(\%)} & \textbf{MACs(G)}                   \\ \hline
DeiT-S(Baseline)                  & 79.80              & 4.54                               \\ \hline
+Token Pruning(\(\rho\)=0.5)             & 78.50 (-1.3)              & 2.11 \\
+W/O inattentive token fusion & 78.40 (-1.4)              & 2.09 \\ \hline 
\end{tabular}
\label{fusion}
\end{table}

\begin{table}[htbp]
\centering
\caption{Comparison of different GELU replacement strategies}
\begin{tabular}{cc}
\hline
\textbf{Method}                                                                                                             & \textbf{ImageNet-1k Top-1(\%)} \\ \hline
Baseline            & 78.40                           \\ \hline
{\textit{\textbf{GELU Replacement Strategies}}} & \\ \hline
\begin{tabular}[c]{@{}c@{}}Full Replacement \\ Train from scratch with ReLU\end{tabular}                                   & 77.00 (-1.4)                          \\ \hline
\begin{tabular}[c]{@{}c@{}}Full Replacement\\ Pretrained (GELU) \& Finetune (ReLU)\end{tabular}                  & 77.92 (-0.48)                          \\ \hline
\begin{tabular}[c]{@{}c@{}}\textbf{Layer-by-Layer Replacement} \\ \textbf{Pretrained (GELU)} \& \textbf{Finetune (ReLU)}\end{tabular} & \textbf{78.36 (-0.04) }                         \\ \hline
\end{tabular}
\label{GELU_replacement}
\end{table}

\subsection{Design Comparison}

Table~\ref{design_comparison} compares the proposed design with previous works. 
The implemented design consists of 496.4K NAND gates and 232KB of SRAM for executing the ViT model. Both weights and activations are represented in 8-bit precision. The overall chip area is 1.19mm\begin{math}^2\end{math}, operating at a clock frequency of 1GHz. The power consumption is kept to a modest 443.2mW.

Compared to other designs, our architecture exhibits strong performance in both energy and area efficiency in an end-to-end model execution. Other works except~\cite{FACT} focused on attention only. Although the design presented in~\cite{28nm} achieves superior energy efficiency, it designs for attention only and comes at the cost of significant area overhead due to the inclusion of complex control mechanisms for power-saving predictions. As a result, it does not provide a combined advantage in both area and power efficiency over our design. Similarly, while the work in~\cite{A3} demonstrates improved energy efficiency, it only supports self-attention computations and lacks comprehensive support for the full transformer block, including QKV generation and FFN. Note that our results and \cite{SpAtten} are based on post-synthesis simulations, while others are from chip measurements and post-layout simulations.

Our approach integrates an algorithm-hardware co-design methodology, effectively reducing computational load and increasing processing speed with minimal overhead. This co-design strategy ensures efficient hardware operation while maintaining a balance between energy efficiency and area constraints, making it a well-rounded solution for ViT model workloads.

\subsection{Ablation Study}
\label{sec:ablation study}

\subsubsection{Dynamic Token Pruning}

Table~\ref{fusion} presents the accuracy and MACs comparison for dynamic token pruning with and without the inattentive token fusion technique used in~\cite{EViT}. Specifically, we removed the fusion mechanism, which combines inattentive tokens, to evaluate its impact on accuracy and computational complexity.

As shown in Table~\ref{fusion}, eliminating inattentive token fusion results in a negligible change in accuracy but leads to a slight reduction in overall computational complexity. This suggests that while token fusion may help preserve some structure within the pruned token set, its absence does not significantly affect performance. The removal also simplifies the hardware implementation by reducing the need for additional computations and memory accesses required for token fusion, thereby making the accelerator more efficient.

The selection of the keep ratio $\rho$ in the Top-$K$ pruning involves a direct trade-off: smaller $\rho$ values yield greater computational savings but can lead to more significant accuracy degradation. Our choice of $\rho$ = 0.5 was empirically determined (as detailed in Table~\ref{fusion} for its impact) to offer a substantial MAC reduction (54\% from token pruning alone before FFN2 pruning) while limiting the initial accuracy drop to approximately 1.4\%. This leaves headroom for additional accuracy loss from FFN2 pruning and INT8 quantization, ensuring that total accuracy degradation remains around 2\% since further reduction to $\rho$ = 0.4 will result in an accuracy drop to 2.7\% according to our simulation.

\subsubsection{GELU Replacement Strategies}

Table~\ref{GELU_replacement} compares different strategies for replacing GELU. Directly replacing GELU with ReLU results in a noticeable drop in accuracy, even when training from scratch. However, the layer-by-layer replacement approach, followed by fine-tuning with ReLU, yields the best results, with the accuracy decreasing by only 0.04\%, which is very close to the original model's performance. This strategy proves to be the most effective, preserving accuracy while successfully replacing GELU with the more hardware-efficient ReLU.

\begin{table}[htbp]
\centering
\caption{Comparison of different FFN2 pruning strategies}
\setlength\tabcolsep{2pt} 
\begin{tabular}{ccc}
\hline
\textbf{Method}         & \textbf{ImageNet-1k Top-1(\%)} & \multicolumn{1}{l}{$W_{FFN2}$ \textbf{Skip(\%)}} \\ \hline
Baseline                & 78.36                           &                                                 \\ \hline
{\textit{\textbf{FFN2 pruning Strategy}}}  &         &                                             \\ \hline
Unified Top-$K$ (50\%)      & 77.01 (-1.35)                          & 50.0                                                    \\ \hline
Unified Threshold=1.5       & 77.65 (-0.71)                          & 61.7                                                    \\ \hline
\textbf{Layer-Adapted Threshold} & \textbf{78.08 (-0.28)}                          & \textbf{59.3  }                                                  \\ \hline
\end{tabular}
\label{tab:FFN2_pruning}
\end{table}

\subsubsection{FFN2 pruning strategies}

\begin{figure}[htbp]
\centering
\includegraphics[height=!,width=0.8\linewidth,keepaspectratio=true]{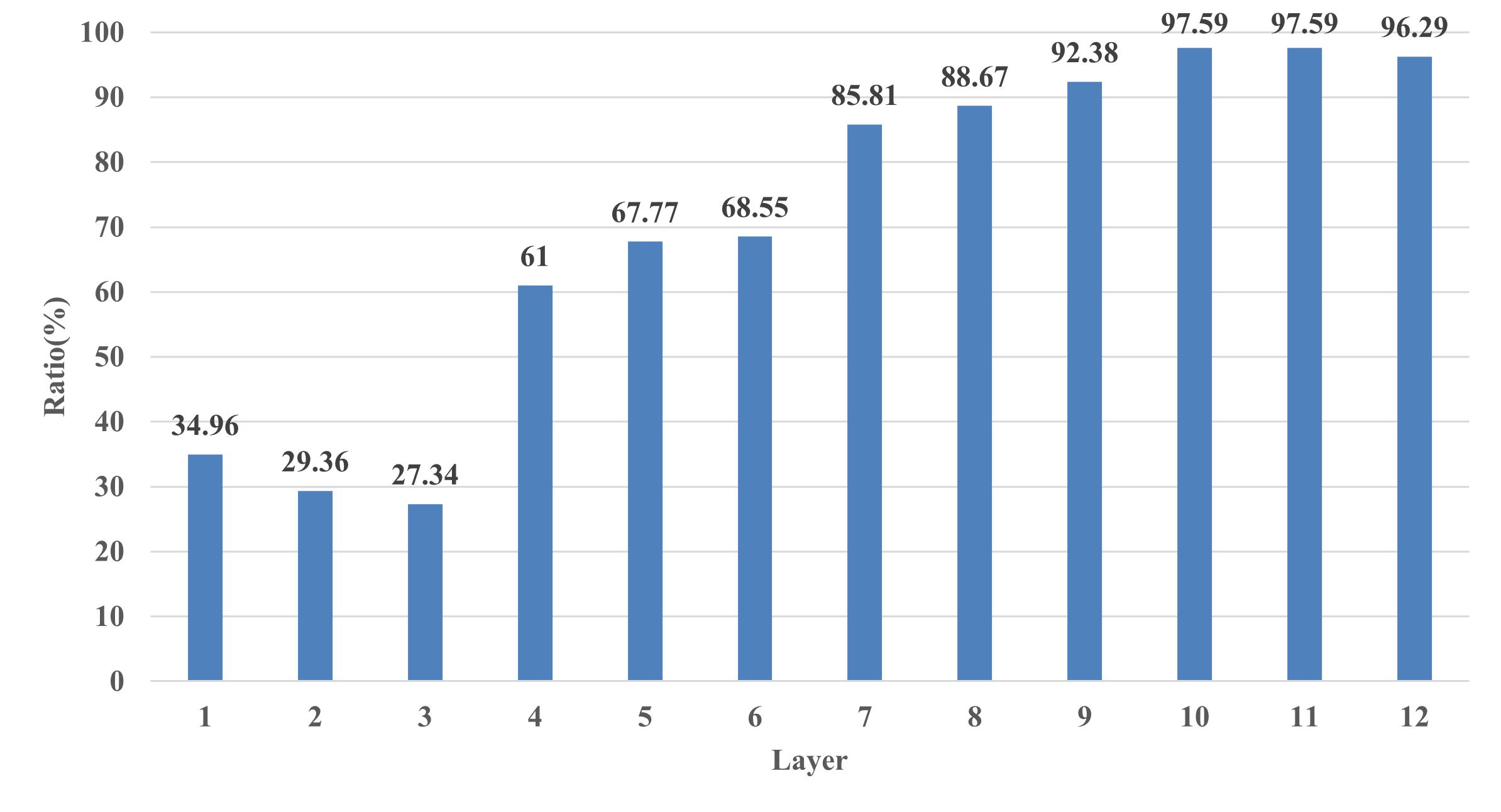}
\caption{FFN2 weight skip ratio across layers.}
\label{ffn2_skip_ratio}
\end{figure}

Table~\ref{tab:FFN2_pruning} compares three FFN2 pruning strategies. For this evaluation, we use the previously fine-tuned model as our baseline and experiment with the following three approaches:

\begin{itemize} \item \textit{Unified Top-$K$ (50\%)}: This approach selects the Top-$K$ most important dimensions for all layers. However, it results in a significant accuracy drop of 1.35\%, making it less than ideal despite its simplicity.
\item \textit{Unified Threshold = 1.5}: This strategy applies a fixed threshold of 1.5 across all layers to determine which dimensions to prune. While it yields a smaller accuracy drop of 0.71\% than the previous one, using a single threshold for all layers may lead to unnecessary pruning, reducing performance.

\item \textit{Layer-Adaptive Thresholds}: This method dynamically adjusts the pruning threshold for each layer based on the distribution of accumulated values. The thresholds across the twelve layers are: [1.5, 1.5, 1.5, 1.0, 1.0, 1.0, 1.0, 1.0, 1.0, 0.5, 0.5, 0.8]. This approach achieves the best balance between accuracy and computational efficiency, with only a 0.28\% drop in accuracy. It also maintains a high skip ratio of approximately 59.3\%. Fig.~\ref{ffn2_skip_ratio} illustrates the FFN2 weight skip ratios across layers, showing that later layers exhibit particularly high pruning rates, reaching up to 97.59\%.
\end{itemize}

\section{Conclusion} \label{chapter:conclusion}

This paper presents a low power vision transformer accelerator with a hardware-software co-design approach, extending optimization beyond self-attention layers to include FFN layers, unlike previous designs. The proposed dynamic token pruning and dynamic FFN2 weight pruning reduce the overall computation by 61.5\%. In addition, replacing GELU with the more hardware-friendly ReLU enhances efficiency. Despite applying these optimizations along with 8-bit quantization, the accuracy loss remains under 2\%.
The proposed pruning methods introduce minimal area overhead while achieving a reduction of 59. 3\% in external memory access for input tokens. The accelerator demonstrates notable improvements in speed and efficiency, achieving 2.31 TOPS/W and 858 GOPS/mm². With a gate count of 496.5K and a 232KB SRAM buffer, the chip occupies 1.19mm\begin{math}^2\end{math}, operates at 1GHz, and delivers a peak throughput of 1024 GOPS, outperforming existing designs in both area and energy efficiency.

\bibliographystyle{IEEEtran}


\begin{IEEEbiography}[{\includegraphics[width=1in,height=1.25in,clip,keepaspectratio]{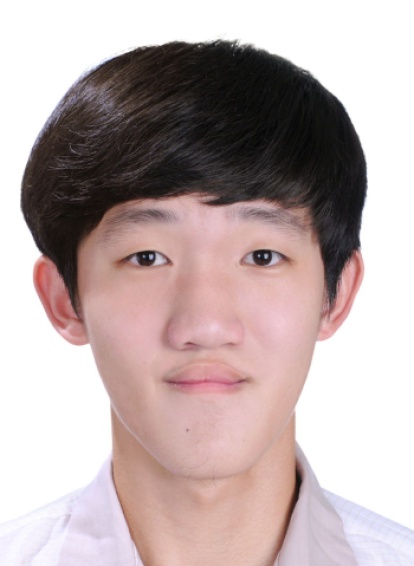}}]{Ching-Lin Hsiung}
received the M.S. degree in electronics engineering from the National Yang Ming Chiao Tung University, Hsinchu, Taiwan, in 2024. He is currently working in the Nvidia, Hsinchu, Taiwan. His research interest includes vision transformer neural network and VLSI design.

\end{IEEEbiography}

\begin{IEEEbiography}[{\includegraphics[width=1in,height=1.25in,clip,keepaspectratio]{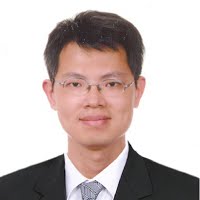}}]{Tian-Sheuan Chang}
	(S’93–M’06–SM’07)
	received the B.S., M.S., and Ph.D. degrees in electronic engineering from National Chiao-Tung University (NCTU), Hsinchu, Taiwan, in 1993, 1995, and 1999, respectively. 
	
	From 2000 to 2004, he was a Deputy Manager with Global Unichip Corporation, Hsinchu, Taiwan. In 2004, he joined the Department of Electronics Engineering, NCTU (as National Yang Ming Chiao Tung University (NYCU) in 2021), where he is currently a Professor. In 2009, he was a visiting scholar in IMEC, Belgium. His current research interests include system-on-a-chip design, VLSI signal processing, and computer architecture.
	
	Dr. Chang has received the Excellent Young Electrical Engineer from Chinese Institute of Electrical Engineering in 2007, and the Outstanding Young Scholar from Taiwan IC Design Society in 2010. He has been actively involved in many international conferences as an organizing committee or technical program committee member.
\end{IEEEbiography}
\end{document}